\documentstyle[aaspp4,psfig,fleqn,11pt]{article}

\newcommand{\ddlnr}[1]{{{{\rm d}#1}\over{{\rm d}\ln r}}}
\newcommand{\dDlnr}[1]{{{\rm d}\over{{\rm d}\ln r}}#1}

\newcommand{\bq}{\begin{equation}}
\newcommand{\eq}{\end{equation}}
\newcommand{\bqa}{\begin{eqnarray}}
\newcommand{\eqaa}{\end{eqnarray*}}
\newcommand{\bqaa}{\begin{eqnarray*}}
\newcommand{\eqa}{\end{eqnarray}}

\newcommand{\Add}{\nabla_{\!\rm ad}}

\lefthead{D. R. Xiong, L. Deng, \& Q. L. Cheng}
\righthead{Nonlocal Time-dependent Convection Theory}

\begin{document}

\title{Turbulent Convection and pulsational Stability of Variable Stars\\
I. Oscillations of Long-Period Variables}
\author{D. R. Xiong$^1$, L. Deng$^2$, \& Q. L. Cheng$^1$}
\affil{$^1$Purple Mountain Observatory, Academia Sinica, Nanjing 210008, P.R. China\\
$^2$Beijing Astronomical Observatory, Academia Sinica, Beijing 100080, P.R. China}

\begin{abstract}
We have performed a linear pulsational stability survey of 6 series of long 
period variable models with  $M=1.0 M_\odot$, $L=3000 - 8000L_\odot$, 
and $(X,Z)=(0.700,0.020),(0.735,0.005)$. The dynamic 
and thermodynamic couplings between convection and oscillations are treated 
by using a statistical theory of nonlocal and time-dependent convection. 
The results show that the fundamental and all the low overtones are always 
pulsationally unstable for the low-temperature models when the 
coupling  between convection and oscillations is ignored. When the coupling 
is considered, there is indeed a "Mira" pulsationally instability 
region outside of the Cepheid instability strip on the H-R diagram. 
The coolest models near the Hayashi track are pulsationally stable. Towards high 
temperature the fundamental mode first becomes unstable, and then the first overtone. 
Some one of the 2nd -4th overtone may become unstable for the hotter models. 
All the modes higher than 4th $(n > 4)$ are pulsationally stable. 
The position and the width of such an instability region on the H-R diagram 
critically depends on the mass, luminosity and metal abundance of the star. 
The overall properties of the dependence are: 
1) For the same mass and luminosity the instability region becomes 
slightly wider and moves to lower effective temperatures as the metal abundance 
increases; 2) For a given chemical abundance, the instability region becomes 
wider and moves to the lower effective temperature as their luminosity 
increases or their mass decreases 
\par
For the luminous red variables seated outside the instability strip the
dynamic coupling between convection and oscillations balances or may even
overtake the thermodynamic coupling. Turbulent viscosity can no longer
be ignored for the pulsational instability of the low-temperature red variables.
The effect of turbulent viscosity becomes more and more important for
higher modes, and may finally become the main damping mechanism of the
pulsation.

\end{abstract}


\section{Introduction}\label{Sect1}
Observationally, there exists a group of low-temperature 
luminous pulsating red variables
to the right of the Cepheid instability strip on the H-R diagram. They are the
most heterogeneous red giants and super-giants belonging to both population I and
II. On GCVS (General Catalogue of Variable Stars), these luminous red variables
are categorized into 3 different types according to their variability: Miras (M),
semi-regulars (SR) and slow irregular variables (L), among which, the first type
(Miras) have been well studied. It is currently believed that the Miras 
are stars on the AGB stage of evolution. 
Detailed summaries of the observational properties and
theoretical work on Miras have been given by Whitelock (1990)
and Wood (1990a,b).
\par
The nonlinear pulsation of Mira variables has been considered by Keeley (1970),
Rose and Smith (1972), Wood (1974), Tuchman, Sack and Barkat (1978,1979), Hill and 
Willson (1979) Bowen (1988) and Perl and Tuchman (1990). The linear analysis is still 
useful for the stability survey. The linear pulsation of Mira variables were 
thoroughly studied by many authors (Kamijo 1962, Langer 1971, Fox and Wood 1982, 
Ostlie and Cox 1986, Balmforth Gough and Merryfield 1990, Cox and Ostlie 1991, Gong, 
Li and Huang 1995), among which Ostlie and Cox (1986) and Gong et al (1995) 
completely ignored the coupling between convection and oscillations. The 
coupling theory used by Kamijo, Langer, Fox and Wood, and Cox and Ostlie is 
over-simplified. Balmforth et al had used Gough's local time-dependent mixing 
length theory of convection in dealing with the coupling.
\par
The red pulsating variables possess very extended convective envelopes. The
convective energy transport in the H and He ionization regions well exceeds
99\% of the total value. Convection overwhelms the $\kappa$-mechanism of radiation
and becomes the principal excitation (damping) mechanism of pulsation.
The local time-dependent theory of convection has been used to interpret the red edge
of the Cepheid instability strip (Baker and Gough 1979, Xiong 1980). When the surface
temperature decreases, the dynamical coupling (through turbulent pressure and
turbulent viscosity) between convection and oscillations becomes more and more
important (Xiong 1977, Gough 1977, Stellingwerf 1984). For red stars outside
the Cepheid instability strip, the dynamical coupling between convection and
oscillations becomes as powerful as the thermodynamic coupling. Precisely
speaking, the local time-dependent theory of convection can no longer (at
least in a self-consistent way) be used to treat the dynamical coupling.
For this reason, we have developed a nonlocal time-dependent statistical
theory of convection (Xiong 1989). In this paper we assumed that the convection
is quasi-isotropic. Such an assumption excludes turbulent viscosity, which
is anisotropic. We have developed a more precise version of
the nonlocal time-dependent statistical theory of convection, in which the
dynamic equations of the third order correlation functions are derived, and the
anisotropy of turbulent convection is also considered (Xiong, Cheng \& Deng
1997a). We have derived an equation of turbulent viscosity which is very
similar to the Stokes equation of viscous fluid.
\par
Comparing with previous works, the present version stands on a solid
base of hydrodynamics for describing the dynamic behavior of convective
motion. Therefore, it would provide a better approach to both the dynamic and 
thermodynamic couplings between convection and oscillations. Using this 
new theory, in this paper we have performed a linear pulsational stability analysis 
for the luminous red stars outside the Cepheid instability strip. The working 
equations of linear non-adiabatic oscillations are given in Sect~\ref{Sect2}. 
Section~\ref{Sect3} describes numerical results and theoretical arguments. 
A summary and concise conclusions of the present work are given at the end.
\section{Working Equations}\label{Sect2}
We have given a complete presentation of the nonlocal time-dependent theory
of convection and all its mathematical formalism, i.e. the radiation-hydrodynamic
equations of stellar oscillations (Xiong, Cheng \& Deng 1997a).
Among all the 14 equations, 3 deal with fluid motion (the conservation
laws of mass, momentum and energy), 2 describe the radiation field
(radiative transfer and energy conservation of the radiation field), and the other 9
equations are given for the second and the third correlation functions of
turbulent velocity and temperature for turbulent convective motion.
Within the frame work of the Eddington approximation, the departure from radiative
equilibrium has been accurately taken into account. The gas and radiation are
treated separately, and they become coupled through the absorption and emission
processes of matter. In the mean time, careful consideration has been given 
to the anisotropy of turbulence and the inertia of turbulent convection.
The turbulent viscosity, which is nearly identical in form to the Stokes
formula for viscous fluid, is included automatically in the dynamic
equation of the second-order correlation of turbulent velocity. When we adopt a
simplified gradient-type diffusion approximation for the third-order
correlations,  this reduces the number of radiation-hydrodynamic equations
down to 11. We have computed the linear non-adiabatic
radial oscillations of p-modes for a solar model using these simplified
radiation-hydrodynamic equations. The results show that the effects
of departure from radiative equilibrium are negligible only except for very
high order p-mode ($n\geq 25$) (Cheng \& Xiong 1997). In this work, we will 
ignore the departure from radiative equilibrium.
In the context of the present study this is safe, because
the giants and super-giants almost always pulsate at low-order modes. Assuming
that the gas and the radiation field are always in equilibrium, we can treat
them together. In this way, the number of equations is reduced down to 10 and
the pulsational equations are greatly simplified. Under such circumstances,
the linear non-adiabatic equations can be written as
\bq
\ddlnr{y_3} + 3y_3 + \delta y_1 -\alpha y_2 =0,
\label{Eq1}
\eq
\bqa
\lefteqn{\dDlnr{\left[\left( P+\delta\rho x^2\right) y_1 +
       \rho x^2\left( 2y_5 - \alpha y_2\right)\right]}}\nonumber\\
 & &\mbox{} +{1\over r^3}\dDlnr{}\left\{\left[{{GM_r\rho r \alpha V\tau_c}\over 4}
            +{4\over 3}{{i\omega\tau^*_c}\over{1+i\omega\tau^*_c}}
            \left(\rho r^3x^2+{{GM_r\rho r \alpha V\tau_c}\over 4}\right)\right]
            \left( \delta y_1-\alpha y_2+3y_3\right)\right.\nonumber\\
& &\mbox{}\left.+{{4GM_r\rho r \alpha V\tau^*_c}\over{3\left(1+i\omega\tau^*_c\right)}}
            \left[y_9-\left(2+{{r^3\omega^2}\over{GM_r}}
            +{3\over 8}i\omega\tau_c\right)y_3\right]\right\}\nonumber\\
 & &\mbox{} -\left\{{3\over{4r^3}}\dDlnr{\left(GM_r\rho r \alpha V\tau_c\right)}
            +4{{GM_r\rho}\over r}+\rho r^2\omega^2\right\}y_3=0,
\label{Eq2}
\eqa

\bq
\ddlnr{y_2}-\ddlnr{\ln T}\left[\chi_{_P}y_1
         +\left(\chi_{_T}-4\right)y_2-4y_3+y_4\right] =0,
\label{Eq3}
\eq
\bqa
\lefteqn{\dDlnr\left\{ 3y_6+{L_c\over L}\left[\left(\delta+C_{p,P}\right)y_1
        +\left(1-\alpha +C_{p,T}\right)y_2+2y_3+y_9\right]+{L_r\over L}y_4\right\}}
        \nonumber\\
 & &\mbox{}+i\omega{{4\pi r^3\rho C_p T}\over L}
           \left[ 3{x^2\over{C_p T}}y_5
           -\left(\Add+{{\delta x^2}\over{C_p T}}\right)y_1
           +\left( 1+{{\alpha x^2}\over{C_p T}}\right) y_2\right]=0,
\label{Eq4}
\eqa

\bqa
{{\sqrt{3}\pi c_2r^3Px^3}\over{GM_rL}}\ddlnr{y_5}
        -{L_1\over L}\left\{\left( 1+\delta\right)y_1
        -\alpha y_2 +6y_3 +3y_5 - y_6\right\}=0,
\label{Eq5}
\eqa

\bqa
\lefteqn{ {{\sqrt{3}\pi c_2r^3\rho C^2_p P xZ}\over{GM_r}}\ddlnr{y_7}
         -L_3\left\{\left[ 1+2\left( \delta +C_{p,P}\right)\right] y_1
         +2\left(C_{p,T}-\alpha\right) y_2 \right.}
         \nonumber\\
 & &\mbox{} \left. + 6 y_3+y_5+y_7- y_8\right\}=0,
\label{Eq6}
\eqa

\bqa
\lefteqn{ {{\sqrt{3}\pi c_2r^3\rho C_p P xV}\over{GM_r}}\ddlnr{y_9}
         -L_5\left\{\left[ 1+ \delta +C_{p,P}\right] y_1
         +\left( C_{p,T}-\alpha\right) y_2\right.}
         \nonumber\\
 & &\mbox{}+\left. 6 y_3 +y_5 + y_9 - y_{10}\right\}=0,
\label{Eq7}
\eqa

\bqa
\lefteqn{ {1\over L}\ddlnr{\left( L_1 y_6\right)}+
         {{4\pi GM_r\rho r}\over{3L}}\left\{\left[\left(\delta-1-{\delta\over 2}
         i\omega\tau_c\right){{1.56\rho x^3}\over{c_1 P}}
         -\alpha V\left( \alpha_p+{\delta\over 4}i\omega\tau_c\right)\right] y_1 \right.}
\nonumber\\
 & &\mbox{} +\alpha\left[\left( {1\over 2}i\omega\tau_c -1\right)
            {{1.56\rho x^3}\over{c_1 P}}+V\left({1\over 4}i\omega\tau_c
           -\alpha_T\right)\right] y_2
\nonumber\\
 & &\mbox{} +\left[ \alpha V\left(2+{{r^3\omega^2}\over{GM_r}}-{3\over 4}
            i\omega\tau_c\right)-2{{1.56\rho x^3}\over{c_1 P}}\right] y_3
\nonumber\\
 & &\mbox{}\left. 3\left(1+{1\over 2}i\omega\tau_c\right)
            {{1.56\rho x^3}\over{c_1 P}}y_5 -\alpha Vy_9\right\}=0,
\label{Eq8}
\eqa

\bqa
\lefteqn{ {1\over{4\pi r^2\rho^2C^2_p}}\ddlnr{\left(L_3y_8\right)}
         +2V\left\{\ddlnr{y_2}-\Add\ddlnr{y_1}
         +\left(\ddlnr{\ln T}-\Add\ddlnr{\ln P}\right)y_9\right\} }
\nonumber\\
 & &\mbox{} -2\left[{{\delta+C_{p,P}}\over{4\pi r^2\rho^2C_p^2}}\ddlnr{L_3}
            +V\Add\nabla_{\!{\rm ad},P}+{{GM_r\rho Z}\over{c_1 r\rho}}
            \left(x+x_c+\Add \alpha_T\sqrt{3}\eta_ei\omega\tau_cx\right)\right]y_1
\nonumber\\
 & &\mbox{}+2\left[{{\alpha-C_{p,T}}\over{4\pi r^2\rho^2C^2_p}}\ddlnr{L_3}
           +i\omega\tau_c\sqrt{3}\eta_e{{GM_r\rho xZ}\over{c_1rP}}\right] y_2
\nonumber\\
 & &\mbox{} +4\left[\left(\ddlnr{\ln T}-\Add\ddlnr{\ln P}\right) V
           -{{GM_r\rho}\over{c_1rP}}\left( x+x_c\right) Z\right] y_3
\nonumber\\
 & &\mbox{} +{{GM_r\rho Z}\over{c_1 r\rho}}\left\{2xy_5+
            \left[2\left(x+x_c\right)+i\omega\tau_c\sqrt{3}\eta_ex\right]
            y_7\right\}=0,
\label{Eq9}
\eqa

\bqa
\lefteqn{ {1\over{4\pi r^2\rho C_p}}\ddlnr{\left(L_5y_{10}\right)}
         +x^2\left(\ddlnr{y_2}-\Add\ddlnr{y_1}\right)
         +i\omega\tau_c{{\sqrt{3}\eta_eGM_r\rho xV}\over{c_1rP}}\ddlnr{y_3} }
\nonumber\\
 & &\mbox{} -\left[{{\delta+C_{p,P}}\over{4\pi r^2\rho C_p}}\ddlnr{L_5}
         +x^2\Add\nabla_{\!{\rm ad},P}+{{GM_r\rho V}\over{c_1rP}}
         \left(2.56x+x_c+i\omega\tau_c\sqrt{3}\eta_e\Add \alpha_Tx\right)\right]y_1
\nonumber\\
 & &\mbox{}+\left[{{\alpha-C_{p,T}}\over{4\pi r^2\rho C_p}}\ddlnr{L_5}
         +i\omega\tau_c\sqrt{3}\eta_e\left(1-\alpha+C_{p,T}\right)
         {{GM_r\rho xV}\over{c_1 rP}}\right] y_2
\nonumber\\
 & &\mbox{}+\left\{{{GM_r \alpha Z}\over r}\left( 2+{{r^3\omega^2}\over{GM_r}}\right)
         +{{GM_r\rho V}\over{c_1rP}}\left[i\omega\tau_c\sqrt{3}\eta_ex
         -2\left( 2.56x+x_c\right)\right]\right.
\nonumber\\
 & &\mbox{}\left.+2x^2\left(\ddlnr{\ln T}
         -\Add\ddlnr{\ln\rho}\right)\right\}y_3
         +\left[ 2x^2\left(\ddlnr{\ln T}-\Add\ddlnr{\ln P}\right)
         +2.56{{GM_r\rho xV}\over{c_1rP}}\right]y_5
\nonumber\\
 & &\mbox{}-{{GM_r}\over{c_1rP}}\alpha Zy_7+{{GM_r\rho V}\over{c_1rP}}
         \left(2.56x+x_c+i\omega\tau_c\sqrt{3}\eta_ex\right) y_9=0,
\label{Eq10}
\eqa
where
\bqa
y_1={{\delta P}\over P},~~~y_2={{\delta T}\over T},~~~y_3={{\delta r}\over r},
~~~y_4={{\delta L_r}\over L_r},~~~y_5={{\delta x}\over x},\nonumber\\
y_6={{\delta L_1}\over L_1},~~~y_7={{\delta Z}\over Z},~~~y_8={{\delta L_3}\over L_3},
~~~y_9={{\delta V}\over V},~~~y_{10}={{\delta L_5}\over L_5}
\label{Eq11}
\eqa
are the relative complex amplitudes of pulsation. $x$, $Z$ and $V$ 
are the auto- and
cross-correlation functions of turbulent velocity and temperature fluctuation
respectively,

\bq
x^2=\sum^3_{i=1}{\overline{w'_iw'^i}}/3,~~~Z={\overline{{T'}^2}}/\bar{T}^2,~~~
\label{Eq12}
V={\overline{w'_rT'}}/T,
\eq
and $\tau^*_c={3\over 16}\tau_c$. $\tau_c$ is the inertial time scale for
turbulent convection, which is
\[
\tau_c={{c_1 Pr^2}\over{\sqrt{3}\eta_e GM_r\rho x}},
\]
while $L_1$, $L_3$ and $L_5$ are the convective variables relevant to the third
correlations. $c_1$ and $c_2$ are respectively the two convective parameters 
relevant to the dissipation and diffusion of turbulence. In our statistical theory 
of turbulent convection (Xiong 1980,1989), $\l_e= c_1Hp$ (Hp is the local pressure
scale height) is the average size of energy-containing eddies. 
$\epsilon=2\eta_e x^3/\l_e$ 
($\eta_e$ is the Heisenberg eddy-coupling constant) is the turbulent disspation 
(Hinze 1975). $\l=c_2Hp$ is the turbulent diffusion 
length scale. The e-folding length of convective overshooting is about 
$1.4\sqrt{c_1c_2}H_P$ (Xiong 1985). $\alpha$ and $\delta$
are the coefficients of expansion and compression for the gas. Variables with a
subscript "P" (or "T") denote the logarithm partial derivative with respect to "P"
(or "T"), such as:
\[
\chi_{_T}=\left ( {{\partial {\rm ln} \chi} \over{\partial {\rm ln} T}} \right )_P.
\]
More detailed definitions concerning
the above expressions have been given in our previous work (Cheng \& Xiong 1997). 
One thing that should be made clear is that the gas and radiation are treated as a
whole ($P=P_g+P_r$) in the present work, differing from our previous manner.
Therefore, the contribution by radiation to all the material properties and
to all the thermodynamic quantities, such as $\delta$, $\alpha$, $C_p$ and $\Add$,
has been included.
\par
The boundary conditions at the bottom layer are:

\bq
y_3=0,
\label{Eq13}
\eq

\bq
y_2=\Add y_1,
\label{Eq14}
\eq

\bq
y_6=\sqrt{{{12\sqrt{3}\eta_e}\over{1+2\sqrt{3}\eta_e}}
      \left( 2+i\omega\tau_c\right)}y_5,
\label{Eq15}
\eq

\bq
y_8=\sqrt{{{12\sqrt{3}\eta_e}\over{1+2\sqrt{3}\eta_e}}
      \left({2\over\sqrt{3}\eta_e}+i\omega\tau_c\right)}y_7,
\label{Eq16}
\eq

\bq
y_{10}=\sqrt{{{12\sqrt{3}\eta_e}\over{1+2\sqrt{3}\eta_e}}
      \left(2+{1\over\sqrt{3}\eta_e}+i\omega\tau_c\right)}y_9,
\label{Eq17}
\eq

and the boundary conditions at the surface are,

\bq
y_3=1,
\label{Eq18}
\eq

\bq
y_1+\left( 4+{{r^3\omega^2}\over{GM_r}}\right)y_3=0,
\label{Eq19}
\eq

\bq
4y_2+2y_3-y_4=0,
\label{Eq20}
\eq

\bq
y_6=\sqrt{{{12\sqrt{3}\eta_e}\over{1+2\sqrt{3}\eta_e}}
      \left( 2+i\omega\tau_c\right)}y_5,
\label{Eq21}
\eq

\bq
y_8=\sqrt{{{12\sqrt{3}\eta_e}\over{1+2\sqrt{3}\eta_e}}
      \left({2\over\sqrt{3}\eta_e}+i\omega\tau_c\right)}y_7,
\label{Eq22}
\eq

\bq
y_{10}=\sqrt{{{12\sqrt{3}\eta_e}\over{1+2\sqrt{3}\eta_e}}
      \left(2+{1\over\sqrt{3}\eta_e}+i\omega\tau_c\right)}y_9,
\label{Eq23}
\eq

As we have done in the previous paper (Cheng \& Xiong 1997), the lower 
boundary is set in the convective overshooting zone, and the upper one 
is taken to be optically thin enough in
the stellar atmosphere. Explicitly, we have adopted an optical depth of
$\tau=0.01$ for the upper boundary in the present work.
\par
The convective boundary conditions Eqs.~(\ref{Eq15})--(\ref{Eq17}) and
(\ref{Eq21})--(\ref{Eq23}) are given by the asymptotic analysis of the
convective quantities within the overshooting zone. For their details 
we refer to our another work by Xiong et al (1997b), where the spatial 
oscillations of convective variables are amply discussed.

\section{The Numerical Results}\label{Sect3}
We have calculated the linear non-adiabatic pulsation for 6 series of 
models of luminous red stars. Their masses, luminosities, chemical 
compositions and of pulsational instablity region are listed in Table 1.
The convective parameters are $c_1=c_2=0.75$, this leads about the same
energy transport efficiency as the original Vitense theory (Vitense 1958)
when the mixing-length is taken to be one and  half of the local
pressure scale height, i.e. $l=1.5H_P$, and the e-folding length of 
convective overshooting is about $1.4\sqrt{c_1c_2}H_P\approx 1.05H_P$.
The reddest model in the series is very close to the corresponding Hayashi
track, while for the hottest model the convective flux is much smaller
than the radiation flux in the second ionization region of helium.
For these hotter stars, the $\kappa$-mechanism functioning in the second
ionization region of helium has already operated, and it has become 
the main excitation mechanism. Therefore, our models actually cover all the 
possible temperatures of the luminous red stars outside the Cepheid 
instability strip, for the present setting of mass, luminosity and
chemical composition.
\begin{table}
\label{Table1}
\end{table}
\par
In the present work we use a simplified MHD equation of state (Hummer and 
Mihalas 1988, Mihalas et al. 1988, Dappen et al. 1988). The neutral 
helium has been considered as a hydrogen-like atom for convenience of 
calculation of its energy levels. A significant departure only appears 
in the ground and the lower excited states. An analytic approach to the OPAL 
tabular opacities (Rogers and Iglesias 1992) and the low-temperature tabular 
opacities (Alexander 1975) is used for the calculation of opacity.
\subsection{"Mira" Instability Strip}\label{subSect3.1}
Table~2 gives the linear amplitude growth rates of the model series 2 of luminous 
red stars, $\eta=-2\pi\omega_i/\omega_r$, where $\omega=\omega_r+i\omega_i$ is 
the complex angular frequency of linear non-adiabatic oscillations. The first 
column is the serial number of the model, the second is the effective temperature 
$T_e$, the 3rd-7th columns are the amplitude 
growth rates for the fundamental through 4th overtones (taking into account
the coupling between convection and oscillations). We have actually calculated
the first 12 modes ($n=0-11$) for non-adiabatic oscillations.
However, all the modes with $n > 4$ are pulsationally stable for
luminous red stars. For the sake of clarity, we give in Table~2
only the first five modes. Aimed at examining the effect of convection
on the pulsational instability for red stars, we have also computed the 
non-adiabatic pulsation ignoring the coupling between convection and 
oscillations (however, convection has been included in the static models). 
If convection does not vary during the oscillation of a star, the convective 
variables $y_5=y_6=y_7=y_8=y_9=y_{10}=0$, Eqs.~(\ref{Eq5})--(\ref{Eq10}) and the 
convective boundary conditions Eqs.~(\ref{Eq15})--(\ref{Eq17}) and 
(\ref{Eq21})--(\ref{Eq23}) can be omitted. Omitting the terms related to 
convection, 
Eqs.~(\ref{Eq2}) and (\ref{Eq10}) are reduced to 

\bq
{\dDlnr\left(P y_1\right)}
         -\left(4{{GM_r\rho}\over r}+\rho r^2\omega^2\right)y_3=0,
\label{Eq24}
\eq

\bq
{\dDlnr\left({L_r\over L}y_4\right)}
        -i\omega{{4\pi r^3\rho C_p T}\over L}
           \left(\Add y_1 - y_2\right)=0.
\label{Eq25}
\eq
The number of pulsational equations reduces down to 4. That is the decoupling 
between convection and oscillations. Eqs.~(\ref{Eq1}), (\ref{Eq3}) (\ref{Eq24}), 
(\ref{Eq25}) and the boundary conditions Eqs.~(\ref{Eq13}),
(\ref{Eq14}), and (\ref{Eq18})--(\ref{Eq20}) were used to calculate the 
non-adiabatic pulsation of stars ignoring the coupling between convection and 
oscillations. The corresponding growth rates of the first five modes ($n=0-4$) 
are given as the 8--12th columns in Table 2. Depicted in Fig.~\ref{fig1} 
are the variations of the amplitude growth rates of  the fundamental mode and the 
first overtone against the stellar effective temperatures for the 6 model series.
\par
It is clear from Table~2 and Fig.~\ref{fig1} that, when excluding the coupling 
between convection and oscillations, the fundamental mode, the first overtone and 
most of the low order overtone  are pulsationally unstable for all the models of 
luminous red stars. But this is not true. Obviously, this is due to the ignorance
of the coupling between convection and oscillations. More generally speaking,
no good interpretation can be made for the red edge of the Cepheid 
instability strip if one does not take such coupling into consideration
(Iben 1966, Xiong 1980).
\par
When the coupling between convection and oscillations is considered,
most of the low order modes and all the higher order modes ($n > 4$)
become pulsationally stable. However, there does exist a pulsationally unstable
region for the luminous red stars outside the Cepheid instability strip.
The coolest models near the Hayashi track are pulsationally stable. Towards high
temperature the fundamental mode first becomes unstable, and then the first
overtone. Some one of 2nd -4th overtones may become unstable for the hotter models.
All the modes higher than the 4th $(n > 4)$ are pulsationally stable.
For simplicity, and to make it distinct with the other
pulsationally unstable region, we will call it
"Mira instability strip" for the moment, though such a name may not be very
appropriate. This instability strip includes not only the regulars Miras,
but also the semi-regular and irregular variables.
\par
Our calculations show that the pulsational instability of a luminous
red star depends critically on stellar metal abundance, luminosity and mass.
The overall properties are:
\begin{enumerate}
\item For the same mass and luminosity, the instability strip becomes slightly 
wider and its location moves to lower effective temperature in the H-R diagram 
as the metal abundance increases.
\item For the same abundance, the instability strip becomes slightly wider
and its location moves to lower effective temperature in the H-R diagram as
the the luminosity increases or the mass decreases.
\end{enumerate}
\par
The above characteristics of luminous red variables are closely connected with the
structure of the convective zones of these stars. Convection is the dominant
factor that controls the pulsational instability for red stars. ${\rm H_2O}$ and
${\rm TiO}$ contribute most of the absorption in the low temperature region, 
therefore the low temperature opacity strongly depends on the metal abundance.
Moreover, the structure of the stellar convective zone is determined by
the nature of the outermost super-adiabatic layer, hence the extension of
the convective zone depends critically on the metal abundance. That explains
why the Mira instability strip and the Hayashi track move to lower 
temperature as metal abundance increases. In the mean time, the structure of
the convective zone depends also on the mass and luminosity. The above reasons
make it easy to understand the properties of the Mira instability strip.

\subsection{Excitation and Damping of Oscillations}\label{subSect3.2}
For the luminous red stars outside the instability strip, the
convective energy transport exceeds 99\% of the total quantity in the ionization
regions of hydrogen and helium. Fig.~\ref{fig2} and Fig.~\ref{fig3} give the
integrated work $W_P$ versus depth for the luminous stars pulsating at the
fundamental mode.
In Fig.~\ref{fig2}, convection has been included for the static model, while
the coupling between convection and oscillations is ignored in calculations
of non-adiabatic oscillations. The only excitation and damping mechanism
results from radiation.  It can be found from Fig.~\ref{fig2} that for the 
two cooler models the excitation comes mainly from the the outermost gradient 
region of radiation flux, where $\chi_{_T}$ approaches its maximum. 
This excitation mechanism functioning in the gradient region of radiation 
flux will be detailed in our second paper (Xiong, Cheng \& Deng 1997b). The 
radiative excitation arising in the second ionization region of helium 
becomes negligible, because the radiative flux is far smaller than the 
convective flux here. The radiative excitation occurring in the second 
ionization region of helium becomes unnegligible for the hotter model. 
For a clearer display, the variation of $\chi_{_T}$ and 
$\log\left(L_r/L\right)$ in the stellar interior are given in 
Fig.~\ref{fig2} as well. Due to the convective energy transport, the 
radiative damping in the deep interior of a star is
greatly weakened. Hence, the star is always pulsationally unstable
when the coupling between convection and oscillations is ignired.
\par
Fig.~\ref{fig3} plots the quantities for the same luminous red star model,
but with the coupling between convection and oscillations being considered.
$W_P$, $W_{Pt}$ and $W_{vis}$ represent respectively the contributions 
of the gas (and radiation) pressure, turbulent pressure, and turbulent
viscosity. Their exact definitions are referred to our previous work
(Cheng \& Xiong 1997). $W_{all}=W_p+W_{Pt}+W_{vis}$ is the total
integrated work, the value of which at the surface of a star should
be the amplitude growth rate $\eta=-2\pi\omega_i/\omega_r$ given
by the non-adiabatic pulsation calculation. Our results show that
they are very closely matched indeed.
\par
By comparing Figs.~\ref{fig2}~ and \ref{fig3}, one can see that
their characteristics are totally different. It can be found from
Fig.~\ref{fig3} that with consideration of the coupling between convection
and oscillations the excitation (and damping) region of pulsation,
where $W_{all}$ changes obviously, extends deeper into
stellar interior in comparition with Fig. 2. This is surely a result of
the coupling between convection and oscillations.
\par
Convection leads to energy and momentum exchange inside a star, and hence
affects the pulsational instability for variable stars. We call the
effects of convective energy transport on variable stars as
thermodynamic coupling, while the effect of momentum exchange
on variable stars as dynamic coupling.
\par
The gas (and radiation) pressure
term $W_P$ in the total integrated work contains the effect of energy transport 
not only by radiation, but also by convection (i.e. the
contribution by thermodynamic coupling between convection and oscillations).
It is hard to unambiguously separate the effects of convective energy
transport from radiative transport, because the two factors are closely 
coupled in the process of stellar oscillations.
Following the conservation law of energy (momentum), any change in convective
flux (turbulent pressure) must lead to the corresponding change in the 
radiation flux (gas pressure) in order to balance the previous
factor, and visa versa. By the same reason, it is not so easy to separate
unambiguously the contribution of gas (and radiation) pressure $W_P$ from 
that of the turbulent pressure $W_{Pt}$.
\par
The combination of $W_{Pt}$ and $W_{vis}$ is the
dynamic coupling between convection and oscillations. The effect of
turbulent viscosity is to convert the kinetic energy of ordered pulsation
into random turbulent kinetic energy, and this process
happens in the low wave number region of turbulent energy spectra .
Through the nonlinear effects of fluid dynamics, the energy
gained in the pulsational motion is cascaded into higher and
higher wave number regions, and is eventually converted into thermal
energy at the smallest turbulent eddies. Therefore the turbulent
viscosity always works as a damping factor to pulsation, i.e.
$W_{vis}$ is always negative. Although $W_{vis}$ is far less
than $W_P$ and $W_{Pt}$ in absolute values for the fundamental
mode of luminous red stars, we cannot ignore its effect on
the stability of pulsation. The contribution of turbulent viscosity
increases very quickly towards higher modes for luminous red variables
because viscous dissipation is proportional to the square of velocity
gradient which grows as the oscillation mode increases. The turburlent
viscosity becomes the main damping mechanism of the high overtones.
This is the reason why the luminous red stars
become pulsationally stable at high modes. Fig.~\ref{fig4}
demonstrates the integrated work diagram for the first, 3rd and 5th
modes of a $T_e=2550K$ model. Fig.~\ref{fig3}a and Fig.~\ref{fig4}
clearly show the quick increase of $W_{vis}$ towards higher modes.
\par
The coupling between convection and oscillations depends critically
on the ratio of the time-scale of convective motion $\tau_c$ to
that of pulsation $\tau_p=2\pi/\omega_r$, i.e. $\omega_r\tau_c$.
When $\tau_c$ is much smaller than $\tau_p$, i.e. $\omega_r\tau_c\ll 1$,
the convection is in nearly quasi-steady variation. When $\omega_r\tau_c$ 
increases, convection lags behind the pulsational variation. The result 
is that not only its variation amplitude decreases, but also a phase 
lag exists. The interaction between convection and oscillations depends 
on both the amplitude of the convective variations and the phase of such
variations. As we will show later on (Xiong, Cheng \& Deng 1997b), the
$\tau_c$ distribution in the interior of luminous red giants is very 
different from that of the RR Lyrae and the horizontal branch red stars. This 
difference results from the difference of their mass- luminosity ratios 
between them. The long-period variables have very low mass-luminosity ratio. 
Therefore, their internal structure has very high central concentration. This
may explain the difference in pulsational characteristics of these two
kinds of stars. In Figs.~\ref{fig3} and \ref{fig4} we have also given
the curves for $\omega_r\tau_c$ versus depth ($\log T$). This 
shows that, for such luminous red stars,
$\omega_r\tau_c\leq 1$ holds for most of the interior of envelope 
model except the extremely outermost layer.

\section{Conclusions and Discussion}\label{Sect5}
By using a nonlocal and time-dependent theory of convection,
we have very carefully treated the coupling between convection and 
oscillations.  The linear stability analysis for luminous
red stars gives us the following results:
\begin{enumerate}
\item The coupling between convection and oscillations is the 
dominant factor for the pulsational instability. When this coupling is not
considered, all the low-temperature red stars are pulsationally
unstable, while when it is taken into account, a Mira
instability strip will show up outside the Cepheid instability strip;
\item Except for some slightly hotter models which may have pulsationally
unstable second to fourth overtone modes, luminous red variable stars
pulsate in the fundamental mode or the first overtone. All the modes higher
than 4 are pulsationally stable;
\item For the low-temperature red stars, the dynamic coupling between
convection and oscillations is in the same order of magnitude as the
thermodynamic coupling, and may even overtake the later;
\item The effect of turbulent viscosity grows very quickly towards
high overtones for the luminous red stars. For high overtones, it 
becomes the main damping mechanism.
\end{enumerate}
\par
The troublesome spatial oscillations of the thermodynamic quantities
are triggered when the local time-dependent convection theory is used in
calculating the stellar oscillations.
(Keeley 1977, Baker \& Gough 1979, Gonczi \& Osaki 1980). In the present
calculations of the nonlocal convection time-dependent theory the spatial
oscillations still exist but are effectively controlled. For the model of
luminous red variable stars, the spatial oscillations happen at the extreme
outer layer of the atmosphere, where $\omega\tau_c\gg 1$. In the stellar
interiors, where $\omega\tau_c<1$, the troublesome spatial oscillations do
not appear. It is difficult to trace the spatial oscillations 
in Figs.~\ref{fig3} and \ref{fig4}, since the
convective energy transport is far less than the radiative  energy transport,
and the turbulent pressure is far less than the gas  pressure
in the atmosphere of a luminous red giant.
It seems true that the spatial oscillations do not affect the
pulsation instability of luminous red stars to a considerable level.
\par
As for the long disputed pulsational mode for O Ceti, some authors think
that O Ceti is pulsating in the first overtone (Kamijo 1962, Keeley 1970,
Wood 1974 \& 1981, Tuchman, Sack  \& Barkat 1978, Tuchman 1991), while some 
others believe that it oscillates in the fundamental mode (Hill \& Willson 1979,
Willson 1982, Bowen 1988). The theoretical pulsation period
$P$ and pulsation constant $Q$ for a luminous
red star model, with $L=5000L_\odot$, $X=0.700$, $Z=0.020$, 
are given in Table~3. Fig.~\ref{fig5} gives the P and Q of the fundamental 
and first modes versus the effective temperature for 6 model series. Our 
theoretical value of the pulsation constant $Q$ is very close to that of
Ostlie \& Cox (1986), but our $Q_1$ for the first overtone is systematically
smaller than that of Wood (1982) especially at the long period end. Following 
our numerical results, there is no model having its first overtone $Q_1$ that 
exceeds $0.045$ days. Normally, $Q_1\leq 0.041$. Assuming for O Ceti, 
$M\approx M_\odot$, $T_e=2900K$ (Wood 1990a,b) and that it pulsates at the
first overtone, then we have $L\approx 12500L_\odot$ or $M_{bol}\approx -5.5$. 
This would make it $0.7^m$ more luminous than the value of $M_{bol}=-4.8$
given by the P-L relation for Miras in the Large Magellanic
Cloud (Glass et al 1987, Hughes \& Wood 1990). This is unlikely to be
real. Wood (1996b) proposed another picture for O Ceti. If Mira's
P-L relation depends on the metal abundance, then the discrepancy between 
the observed and the theoretical $Q$ values could be removed. 
Our results of linear pulsational stability survey support Wood's guess. 
The instability strip moves indeed towards the low temperature as the metal 
abundance increases, as shown by Table 1 and Fig. 1. 
Our calculations show that if the metal abundance for LMC Miras are
only one-half of that of O Ceti the red edge of the Mira instability strip
in the Galaxy will be lowered by about $350K$ compared with the LMC's. Therefore
the magnitude of O Ceti derived from the LMC P-L relation will be $0.6^m$ more
luminous than it actually is. Let us suppose that the effective temperature
of O Ceti is $T_e=2900K$ and $M_{bol}=-4.2$, then the pulsation period of 
the fundamental mode given by
the non-adiabatic calculation will be 330 days. Following our linear
stability analysis, it appears more likely for O Ceti to pulsate in
the fundamental mode than in the first overtone.
\acknowledgments
This work is supported in part by the contract number 19573018 of
National Natural Science Foundation of China (NSFC).

\newpage\eject
Firure Captions
\figcaption[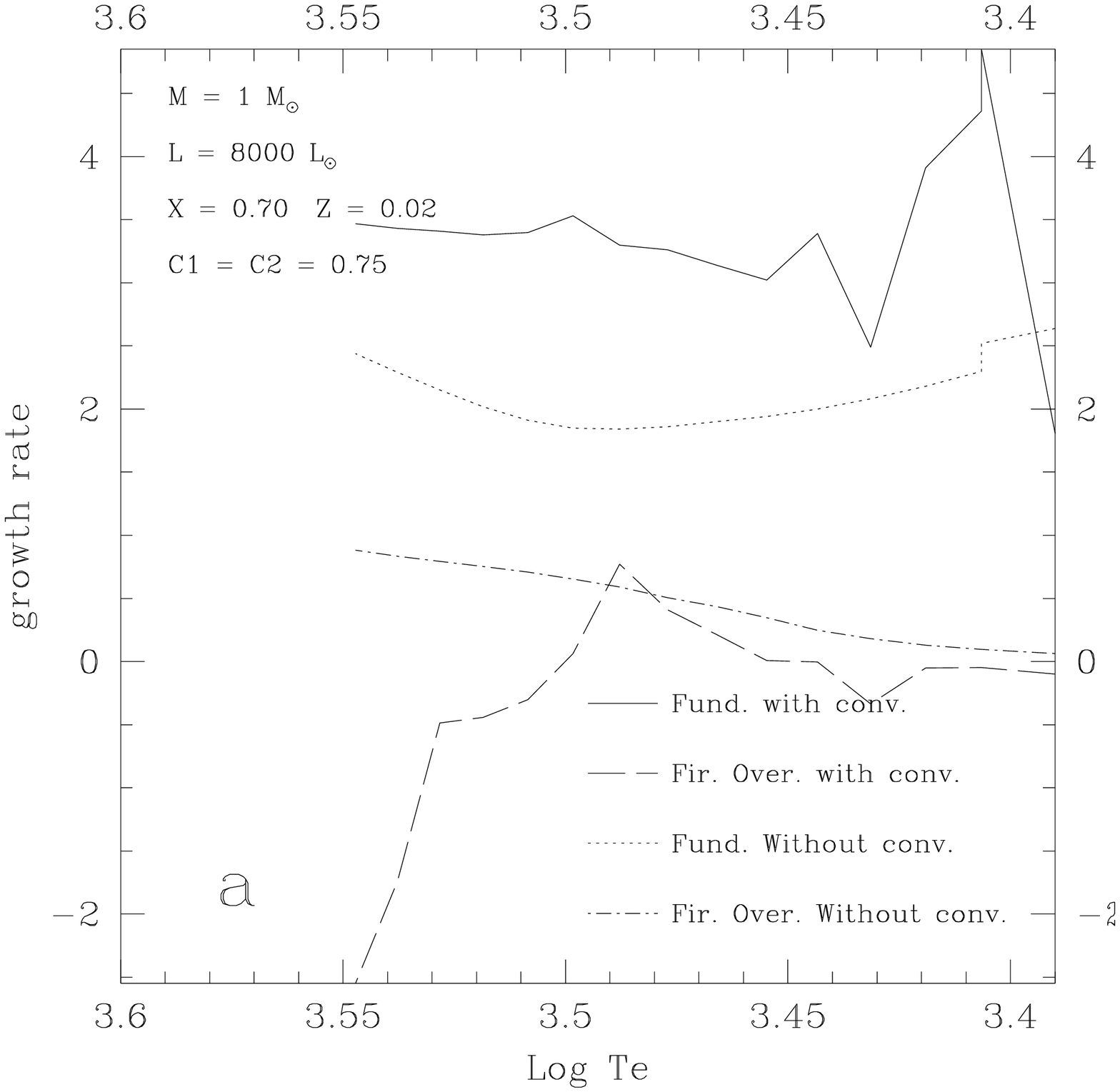,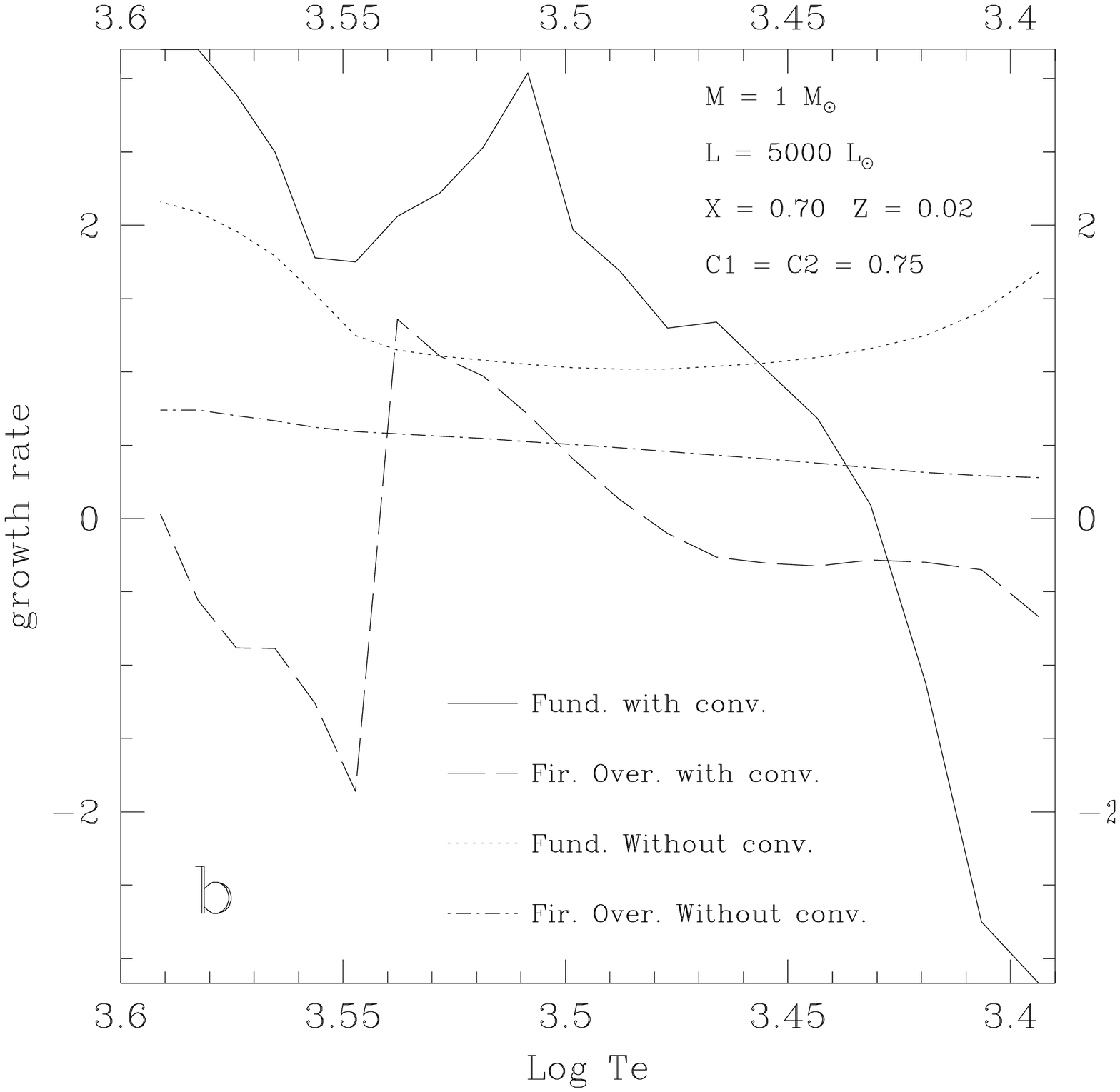,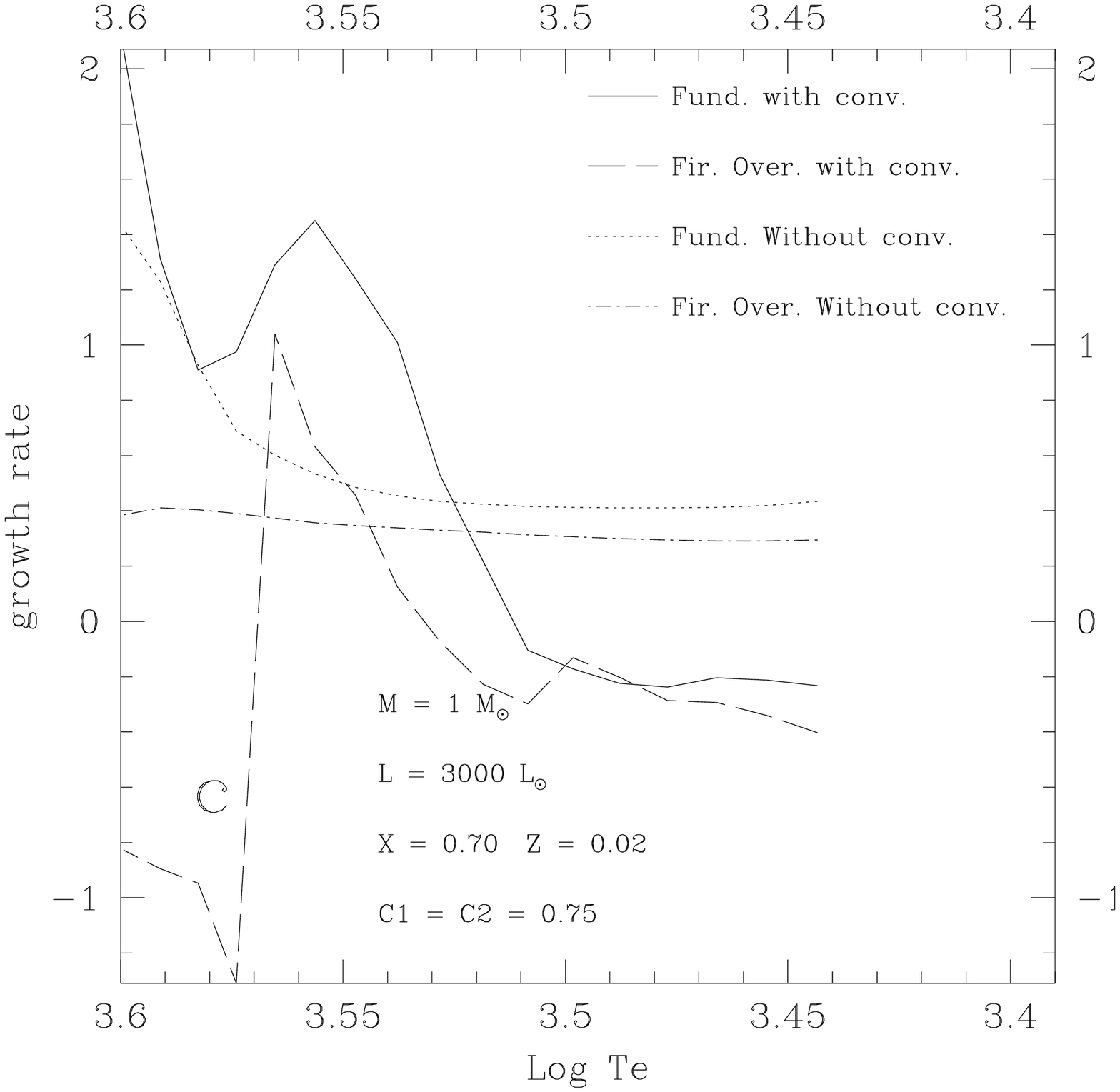,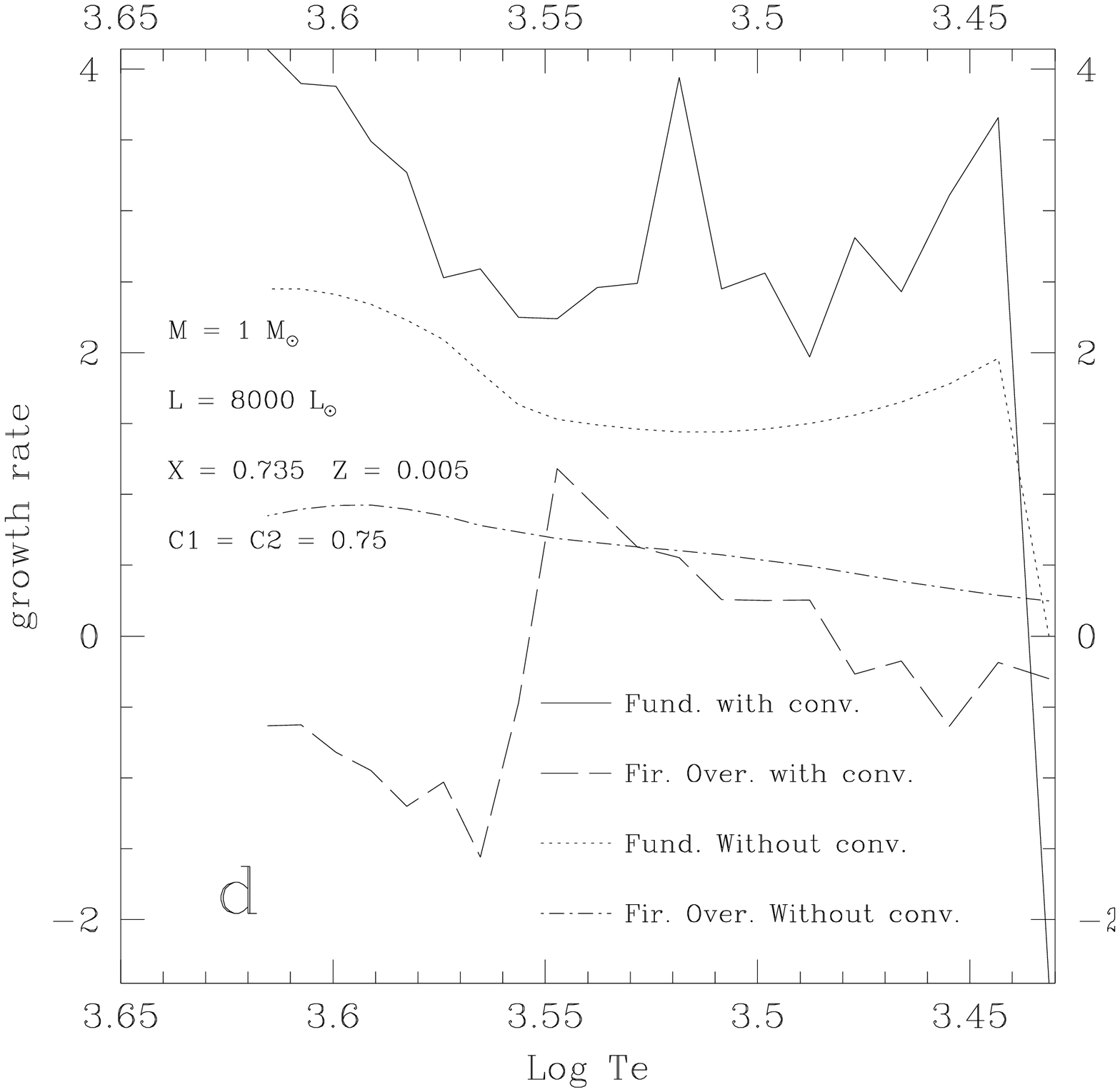,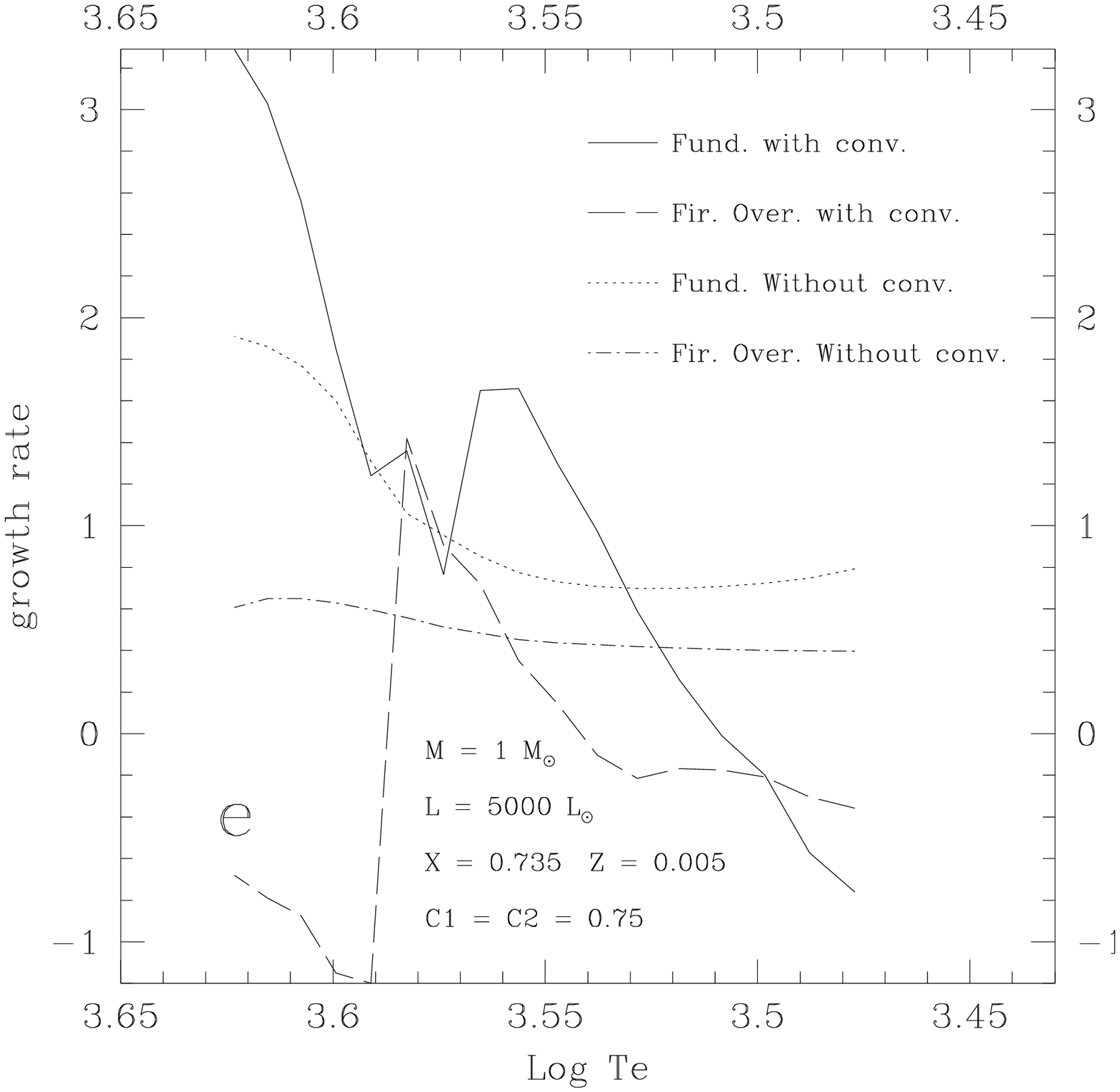,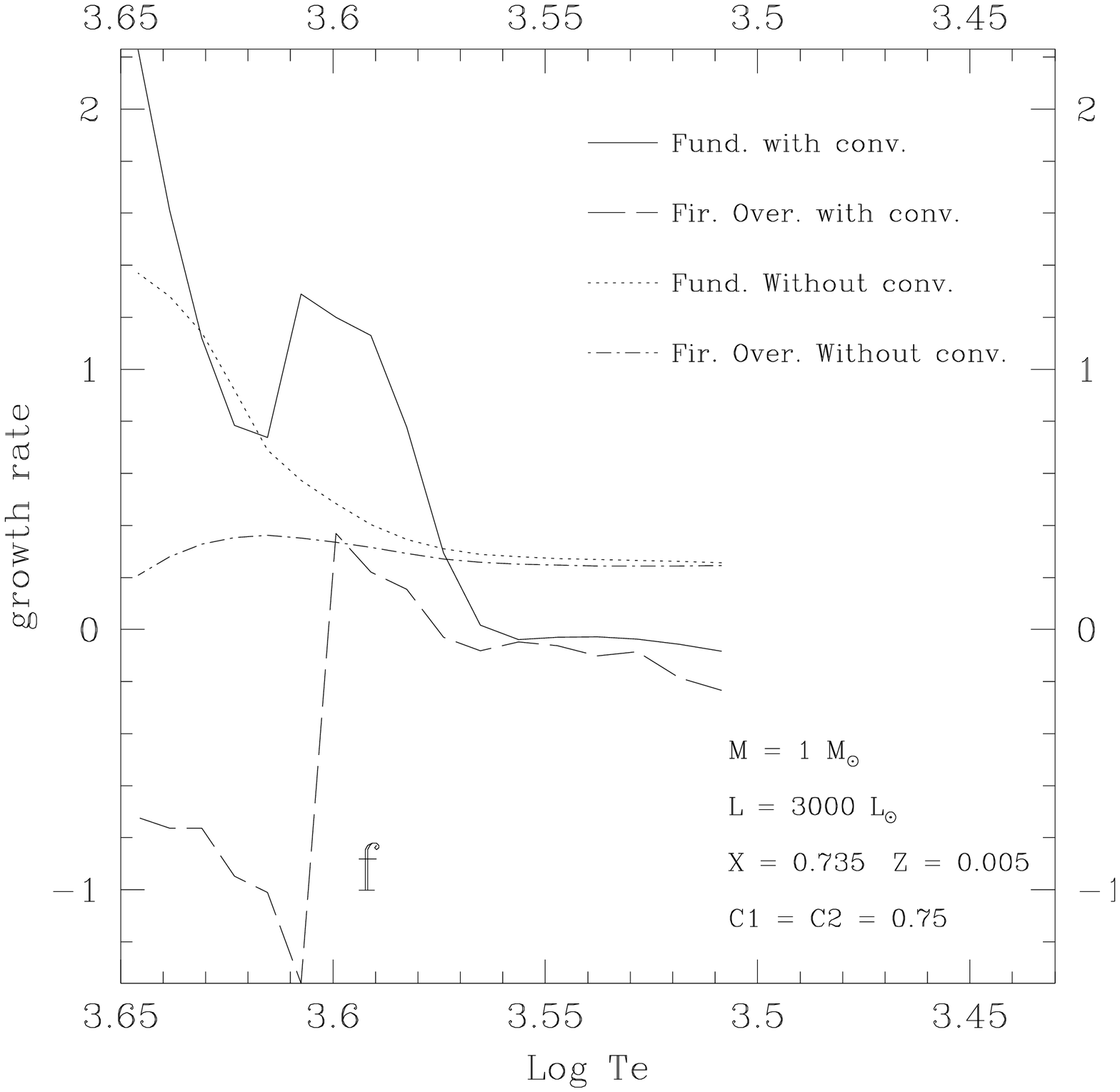]{
The amplitude growth rate of pulsation for the fundamental mode and the first
overtone varies with the effective temperature, {\it with conv.} and 
{\it without conv.} denote the cases when the coupling between convection 
and oscillations are included and not included respectively. 
\label{fig1}}
\figcaption[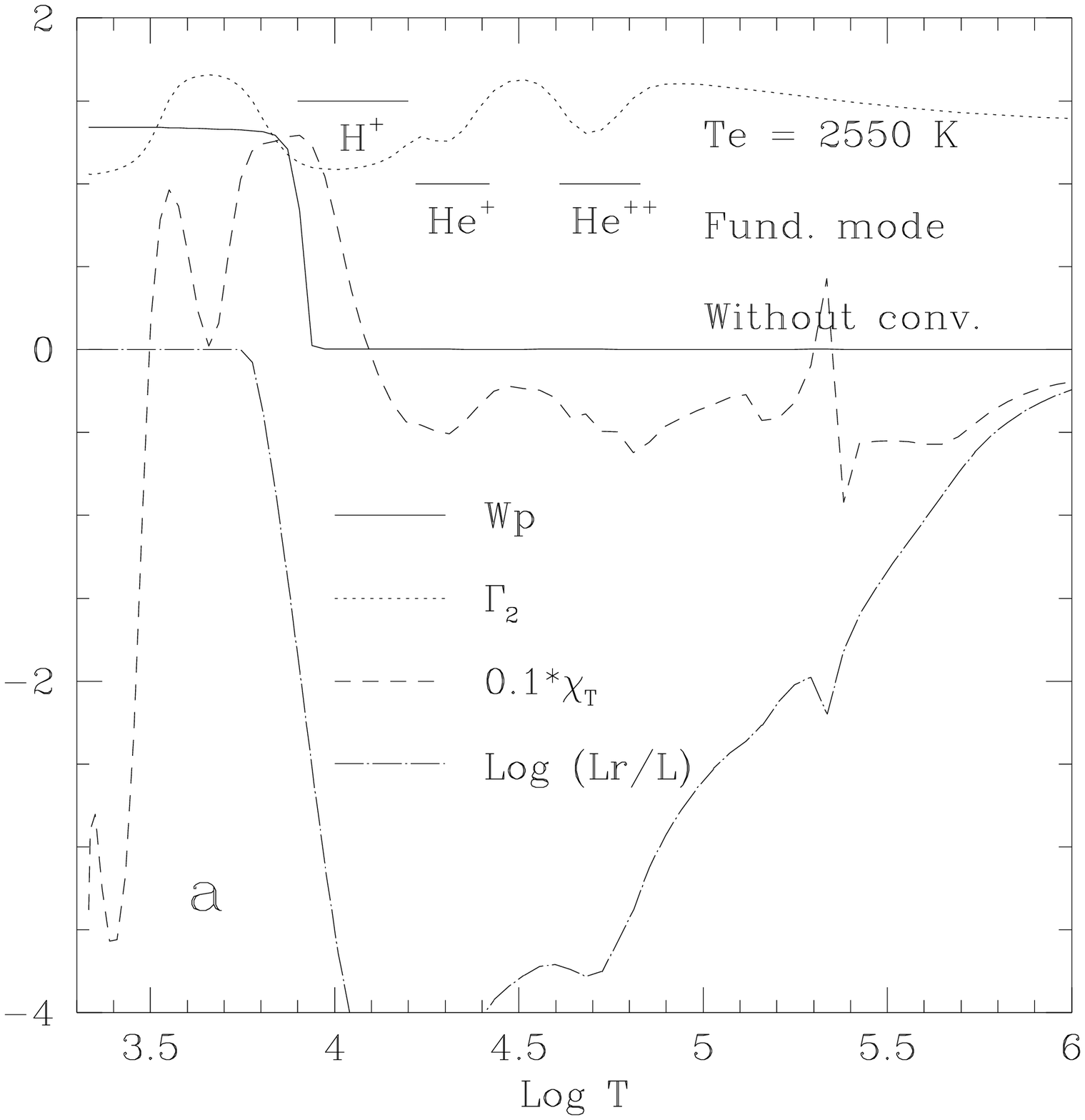,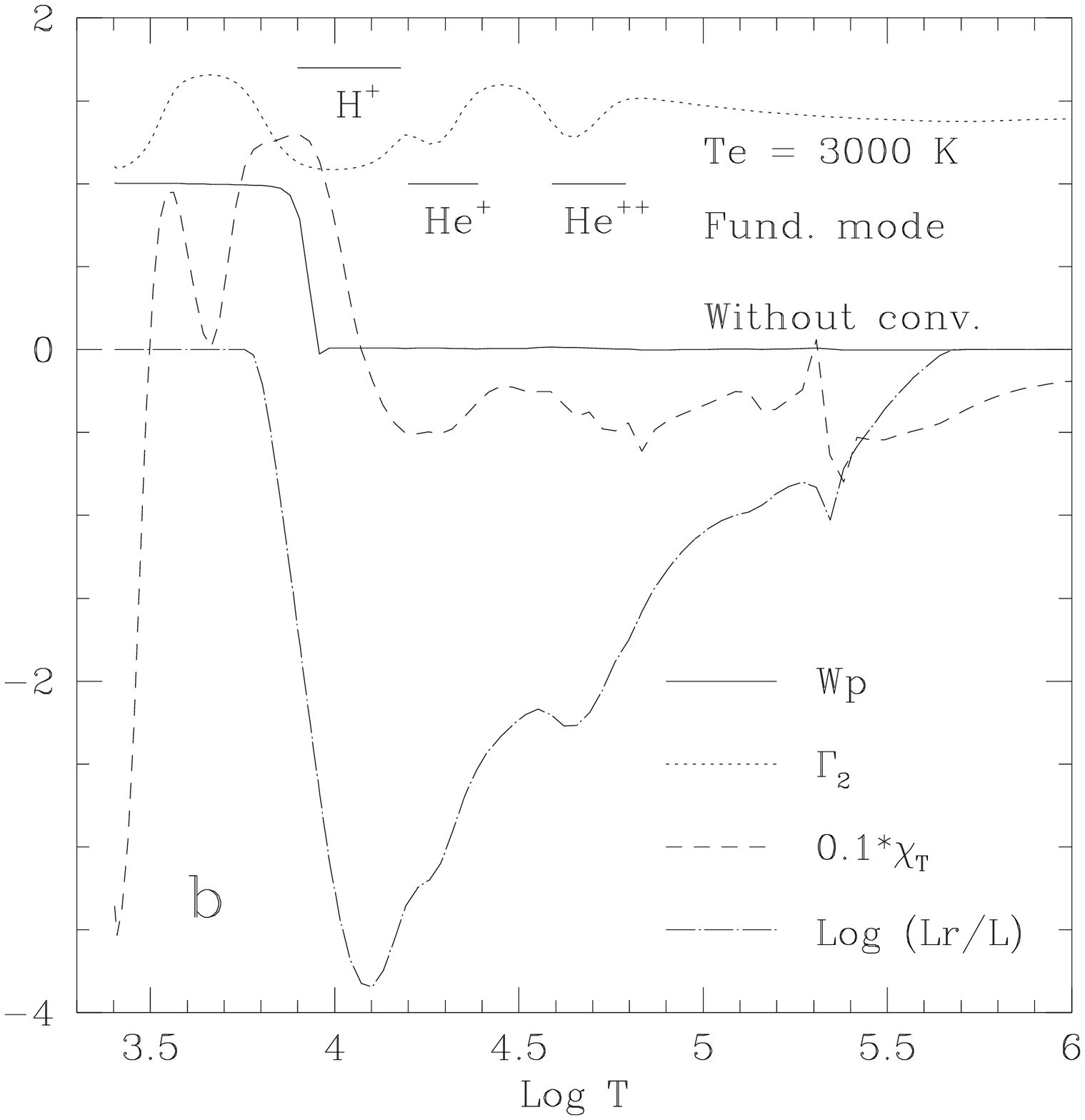,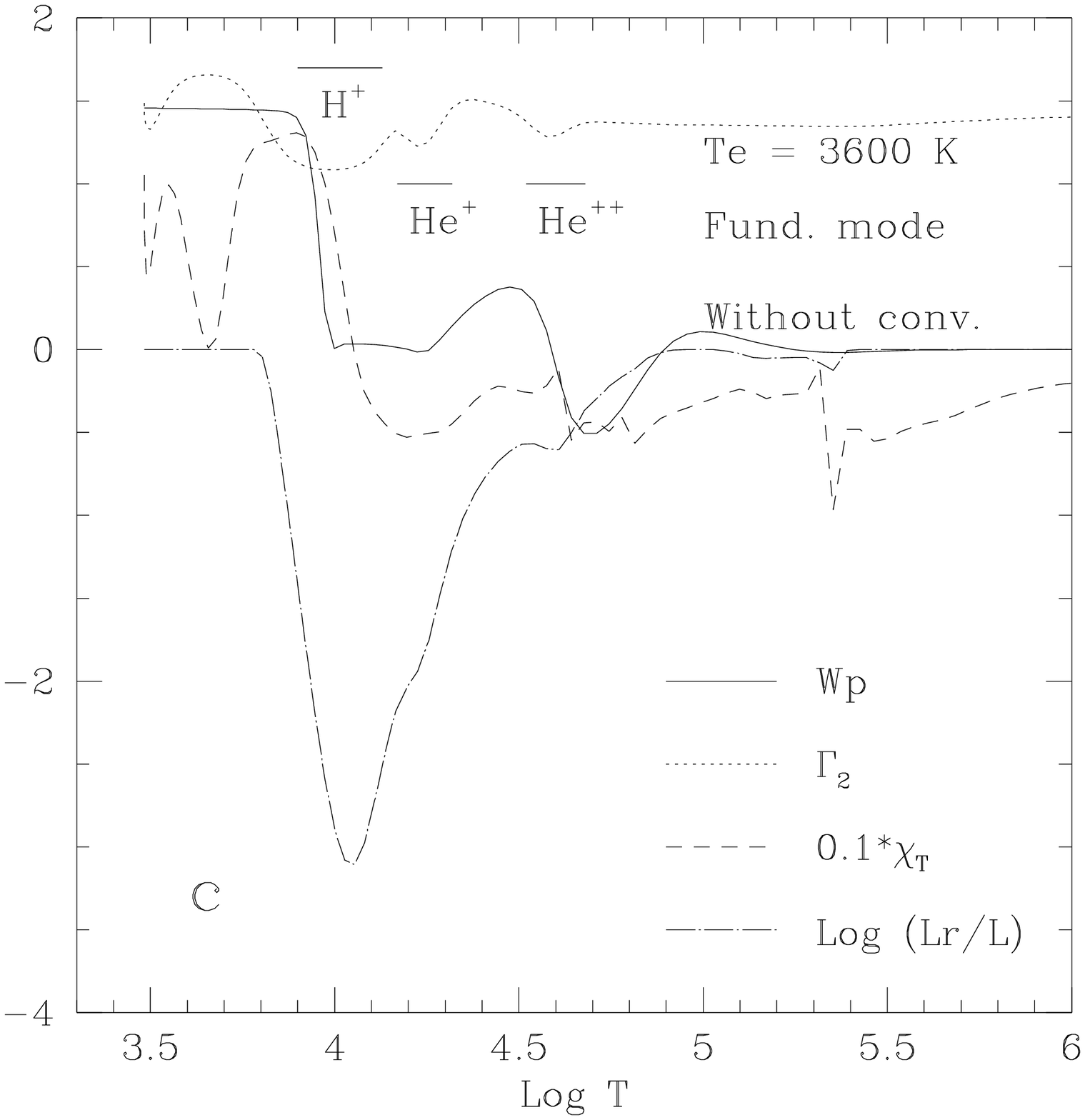]{
The variations of the integrated work $W_p$ (solid line),
$\Gamma_2$ (dotted), $\chi_{_T}$ (dashed), and $\log\left( L_r/L \right)$ with
respect to depth ($\log T$)  for three models of model 
series 2 ($M =5000M_\odot$, $L=5000 L_\odot$, $X=0.700$, $Z=0.020$). The 
coupling between convection and oscillations
is ignored in the pulsation calculation, while convection is included
for the corresponding static models. The three horizontal lines are the
locations of the ionization regions of H and He.
\label{fig2}}

\figcaption[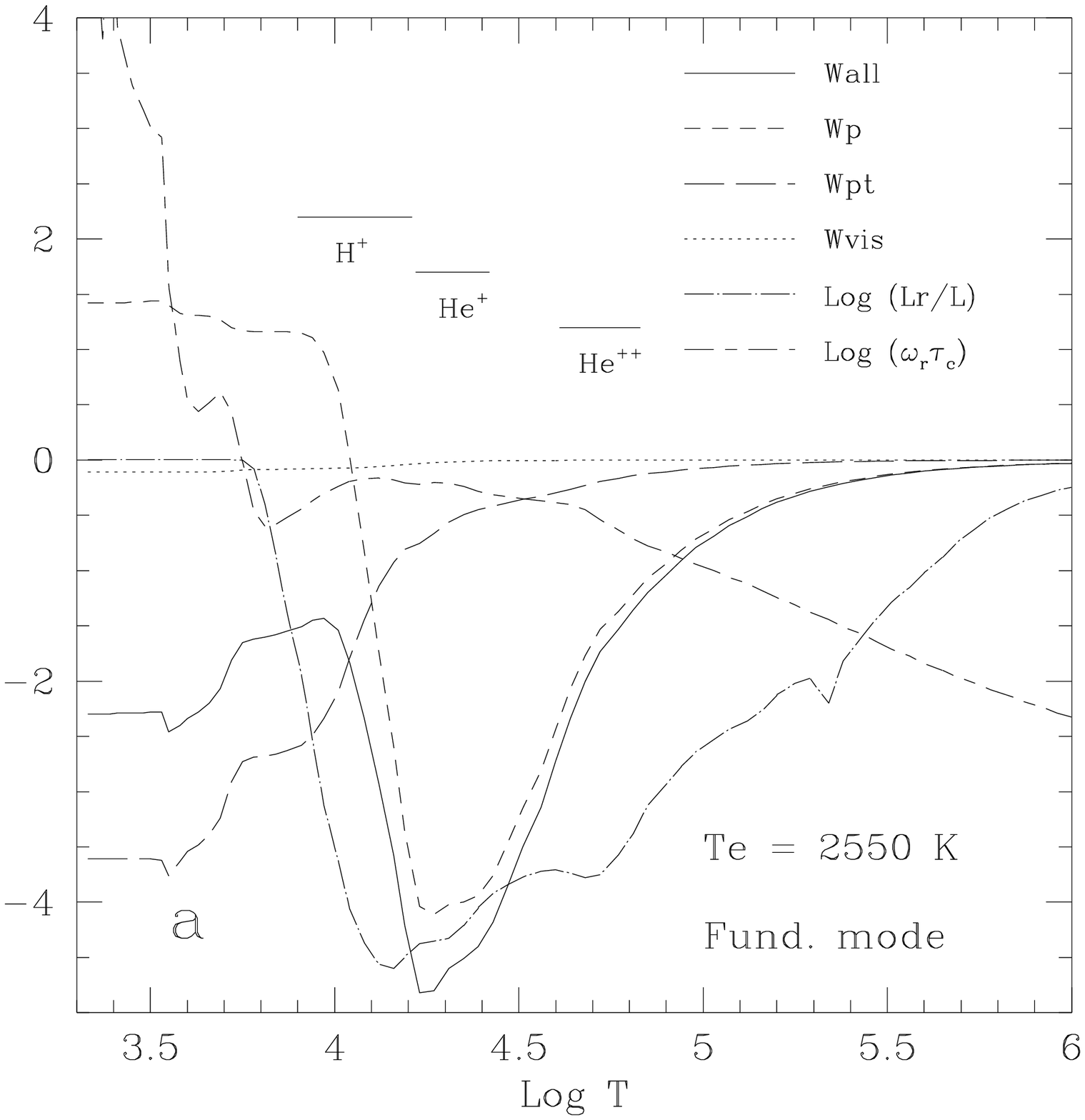,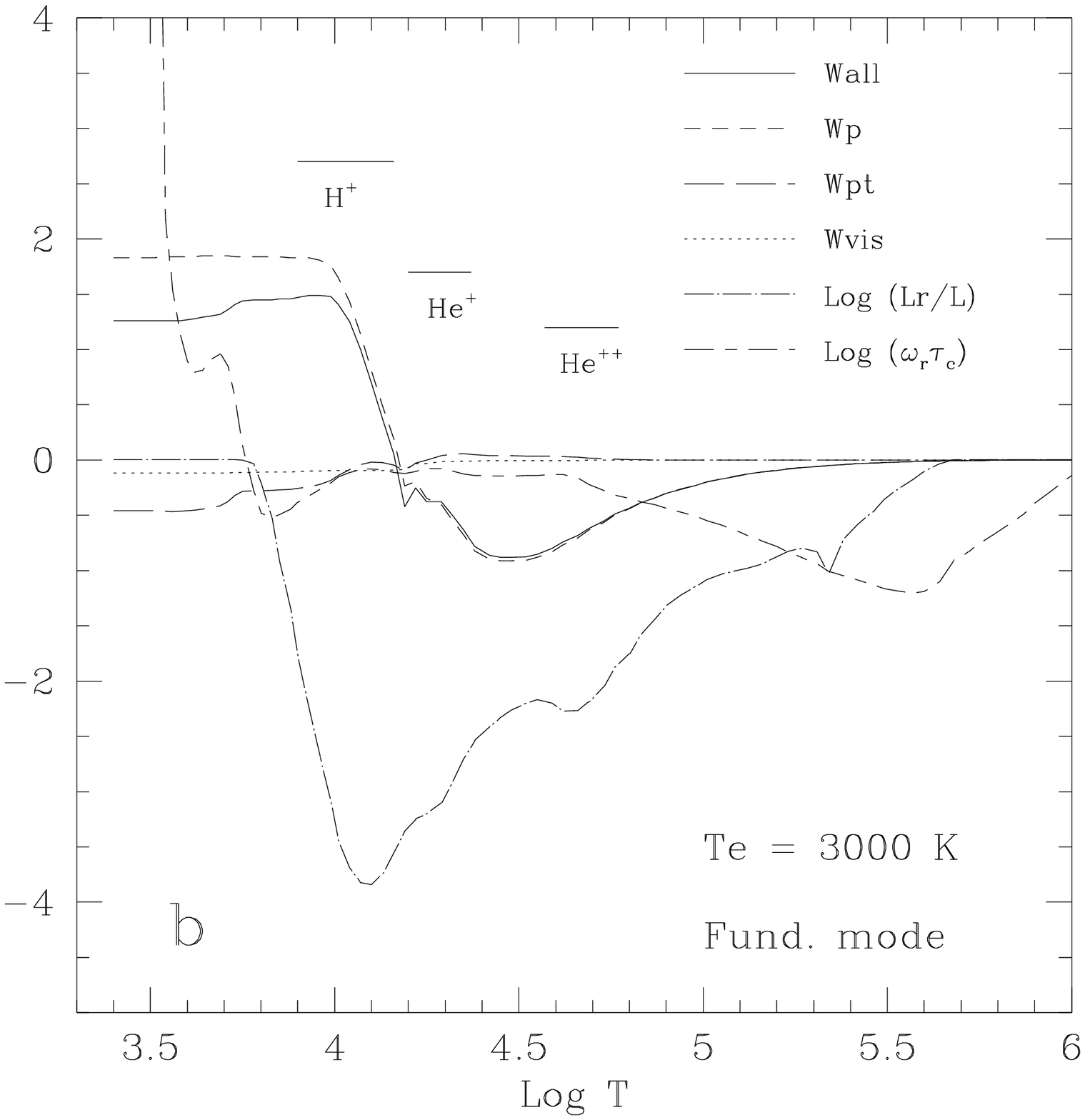,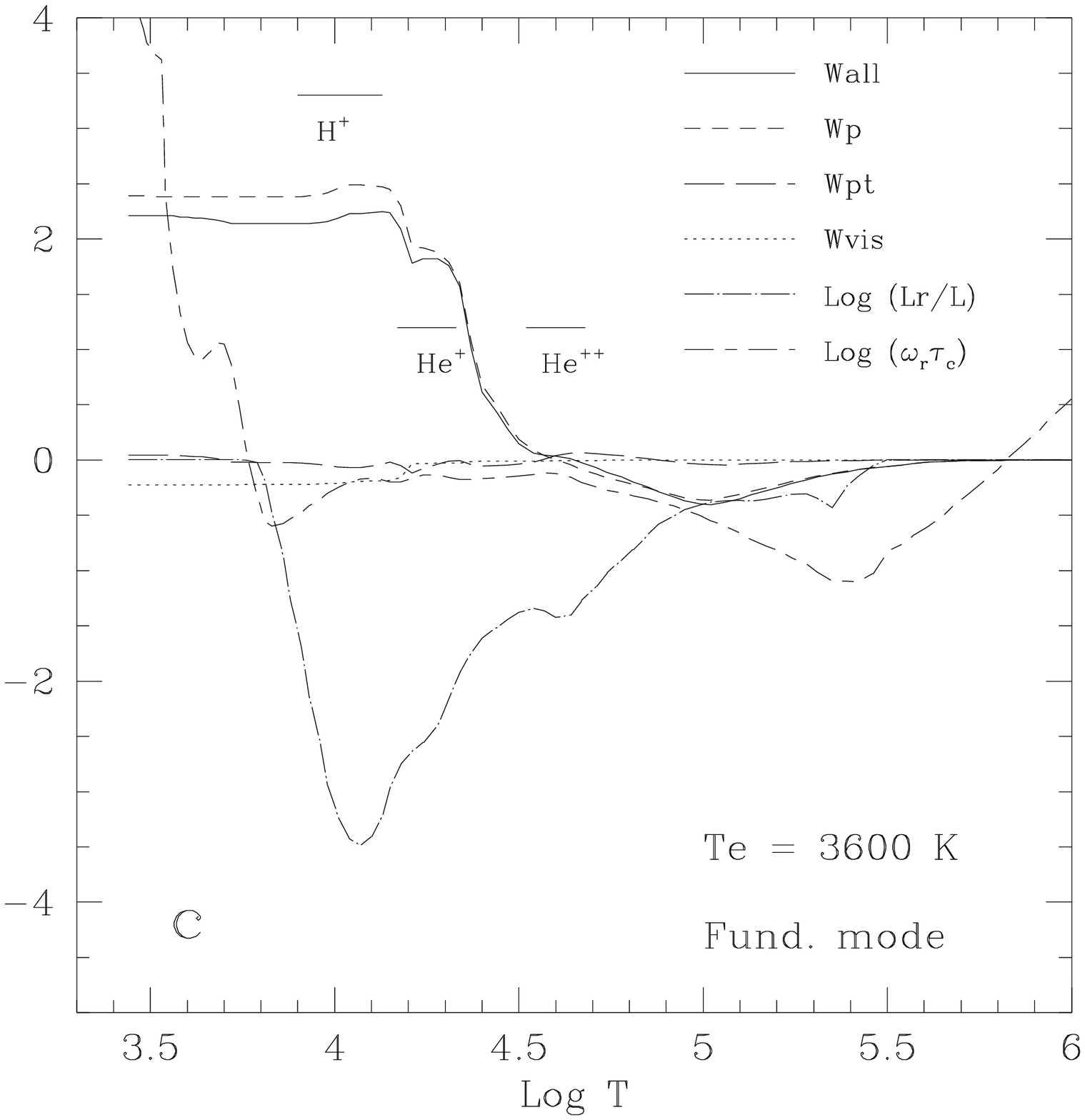]{
The integrated work vs. depth ($\log T$)  for the same models as in Fig. 2,
however the coupling between convection and oscillations is included. 
$W_P$, $W_{Pt}$ and $W_{vis}$ are respectively the components by 
gas (and radiation) pressure, turbulent pressure and turbulent viscosity. 
The variations of $\log\left(L_r/L\right)$ and $\log\left(\omega\tau_c\right)$ 
are also given.
\label{fig3}}

\figcaption[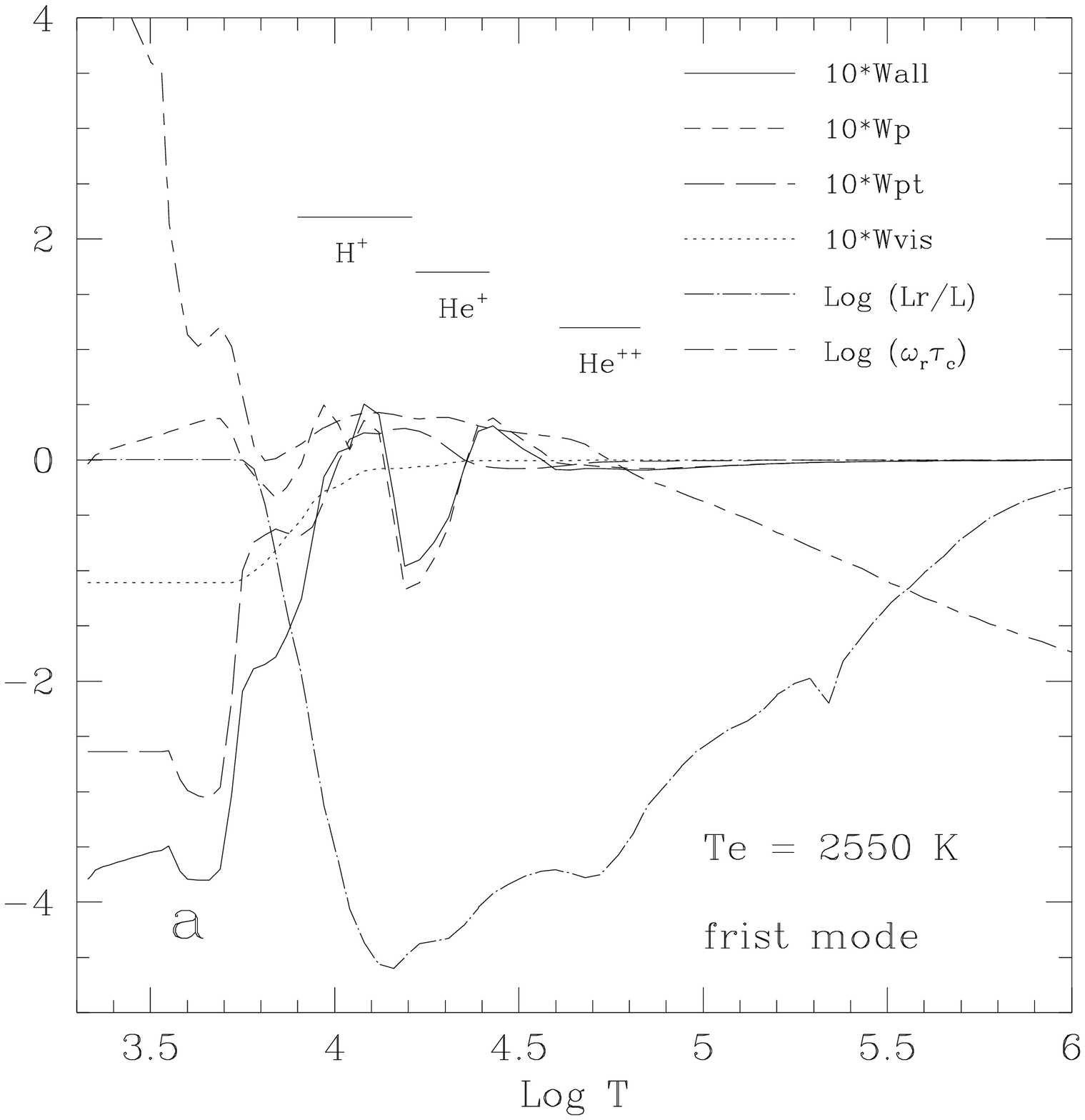,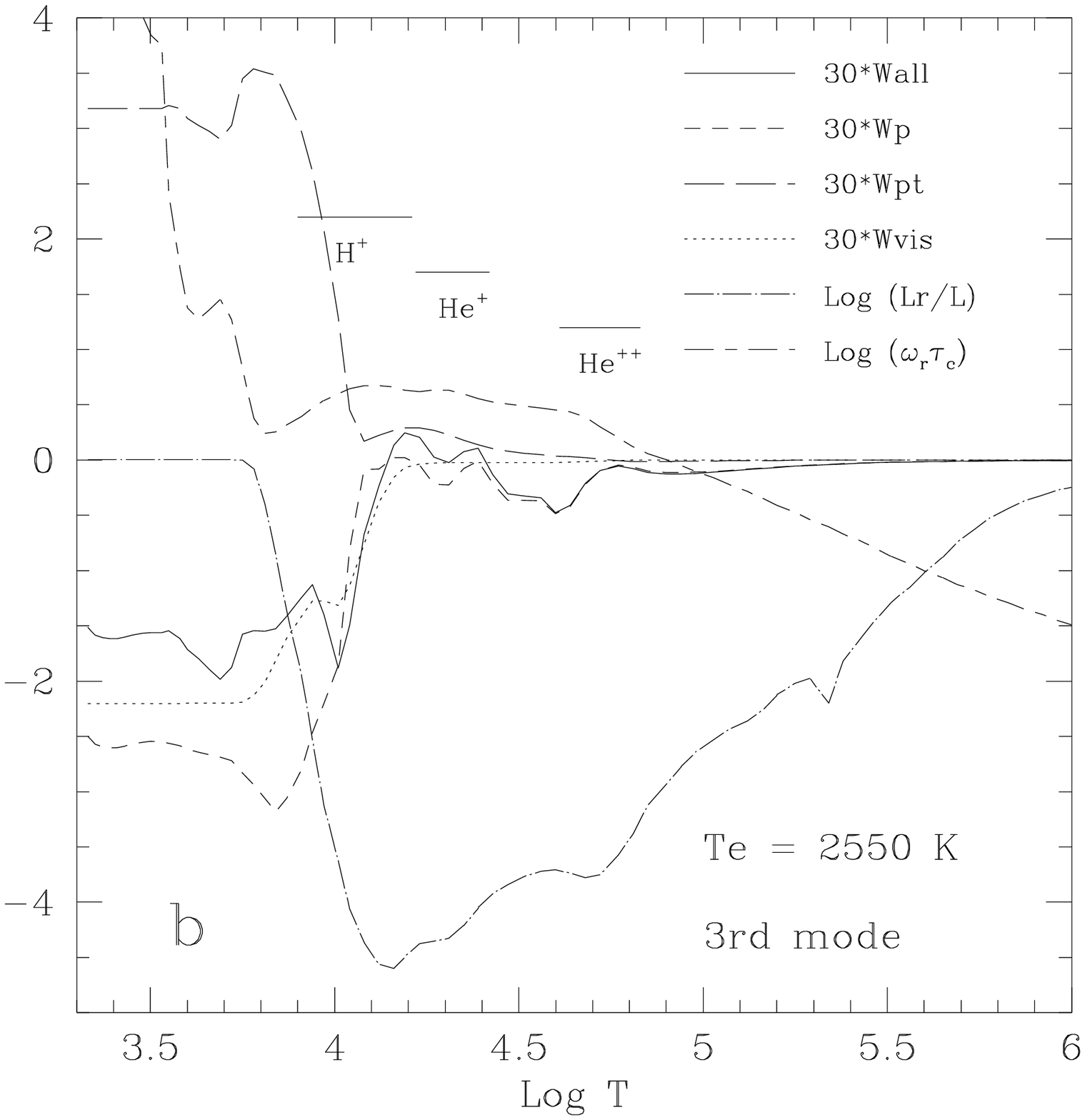,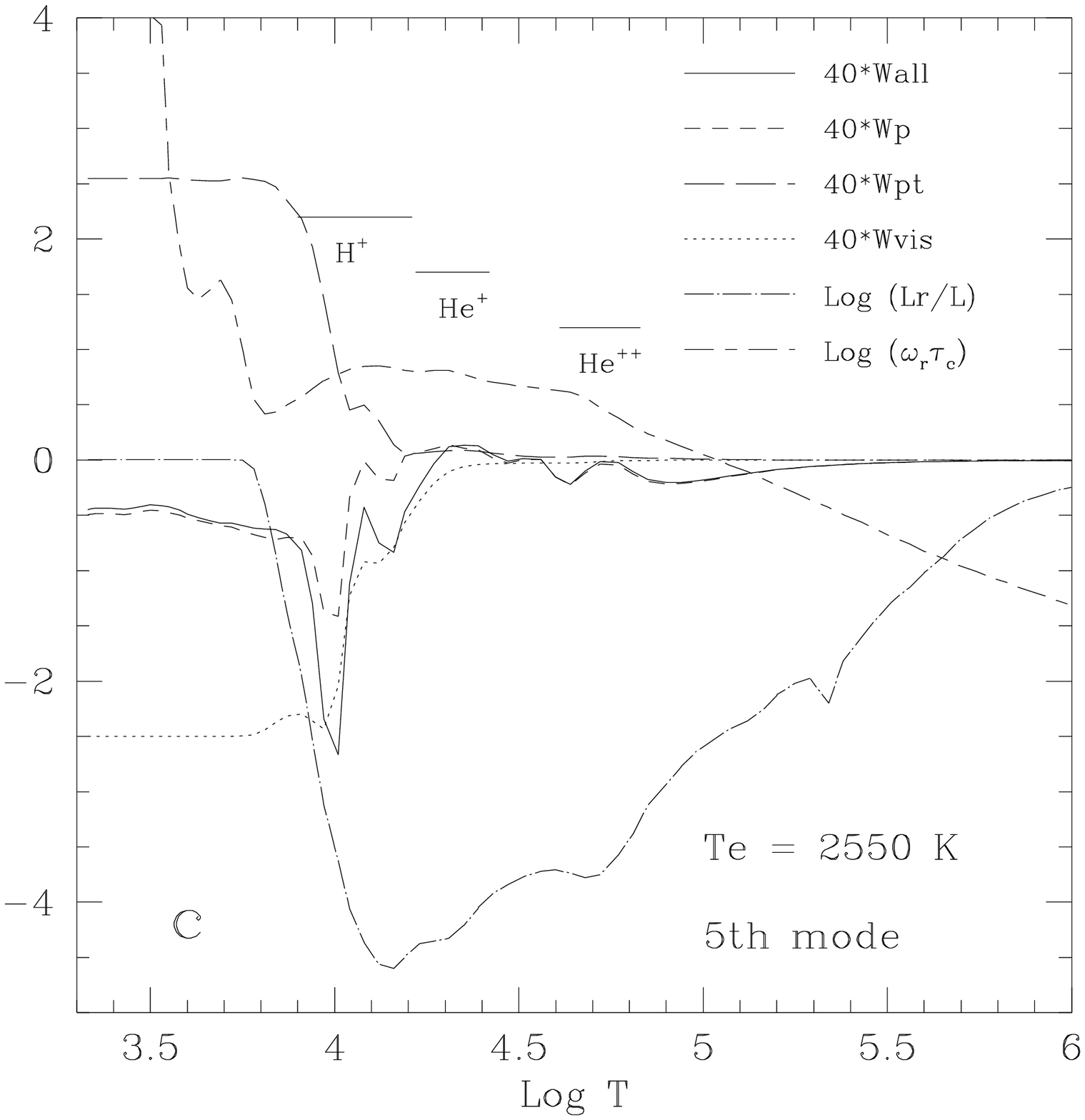]{
The same as Fig.~\ref{fig3}, but for first, third and fifth overtones,
The effective temperature of corresponding static model is $T_e=2550K$.
\label{fig4}}

\figcaption[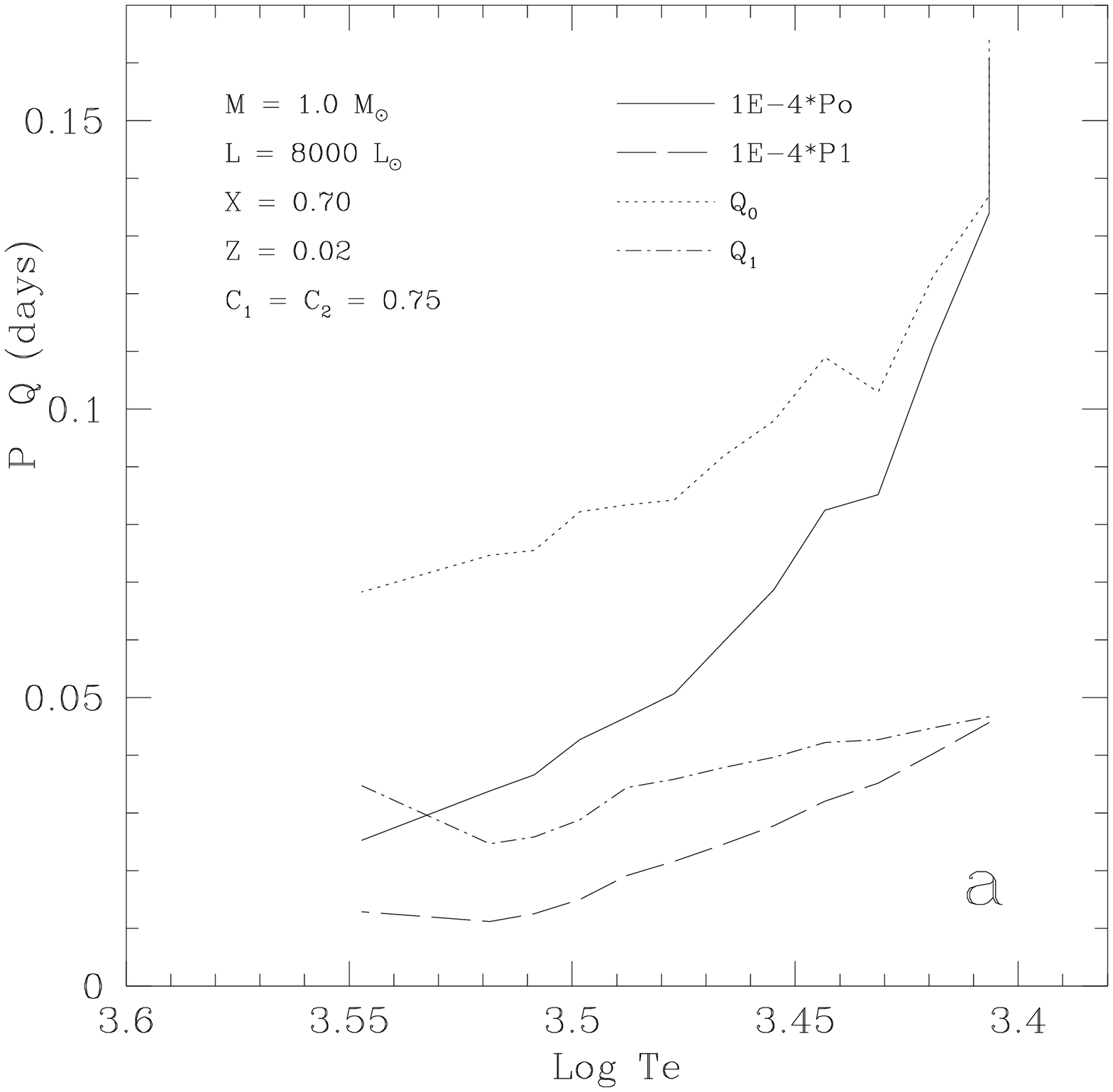,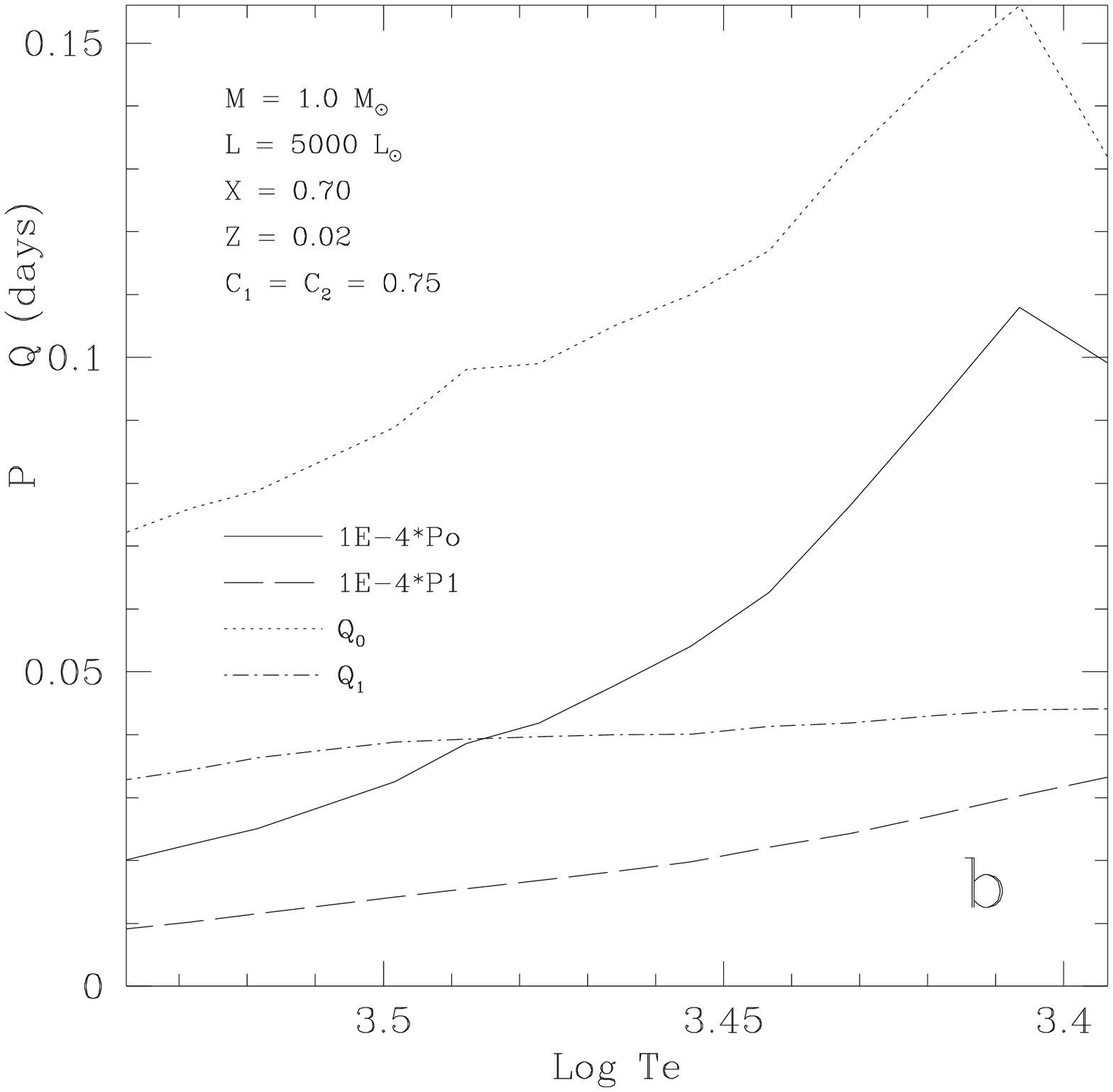,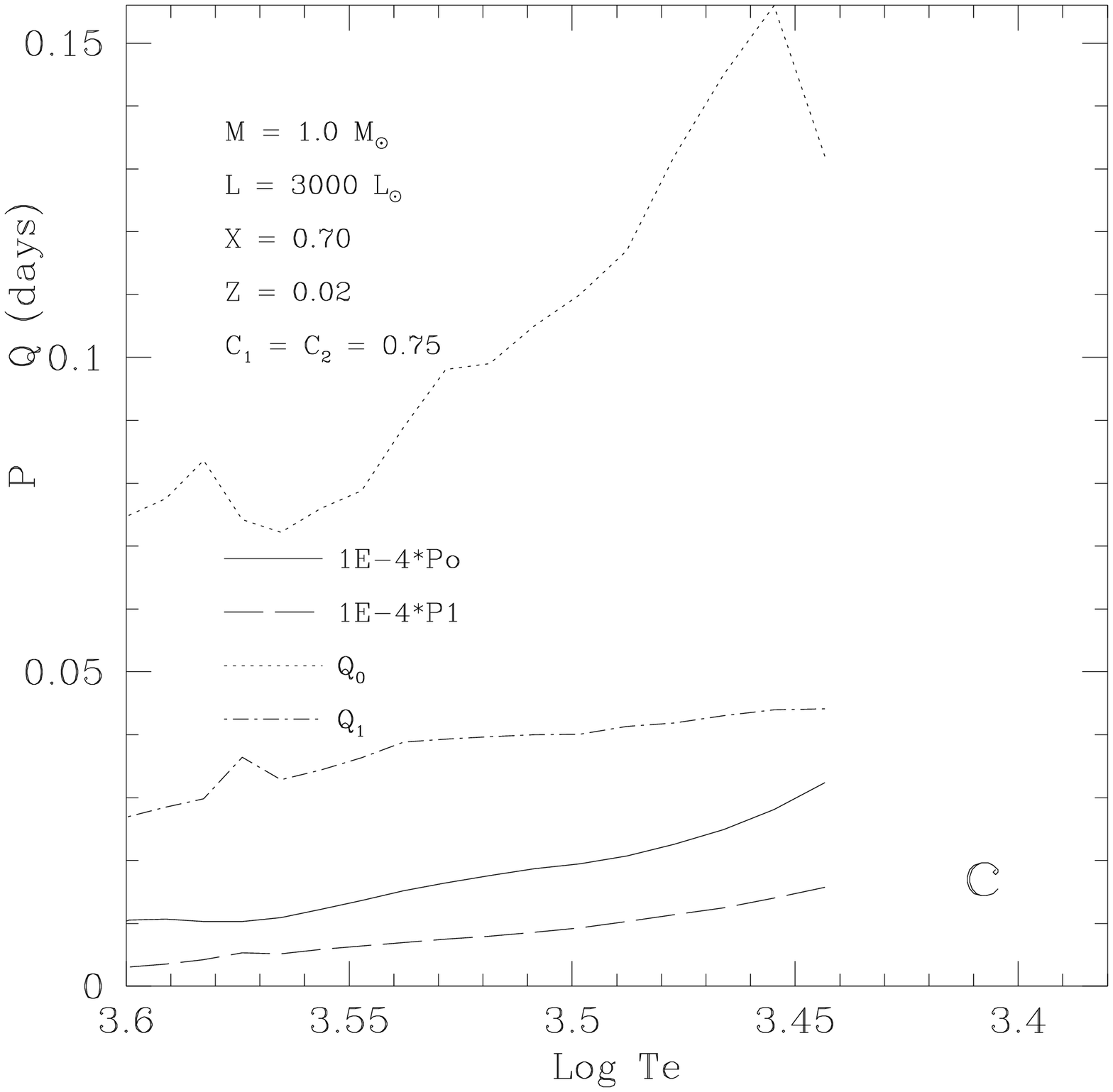,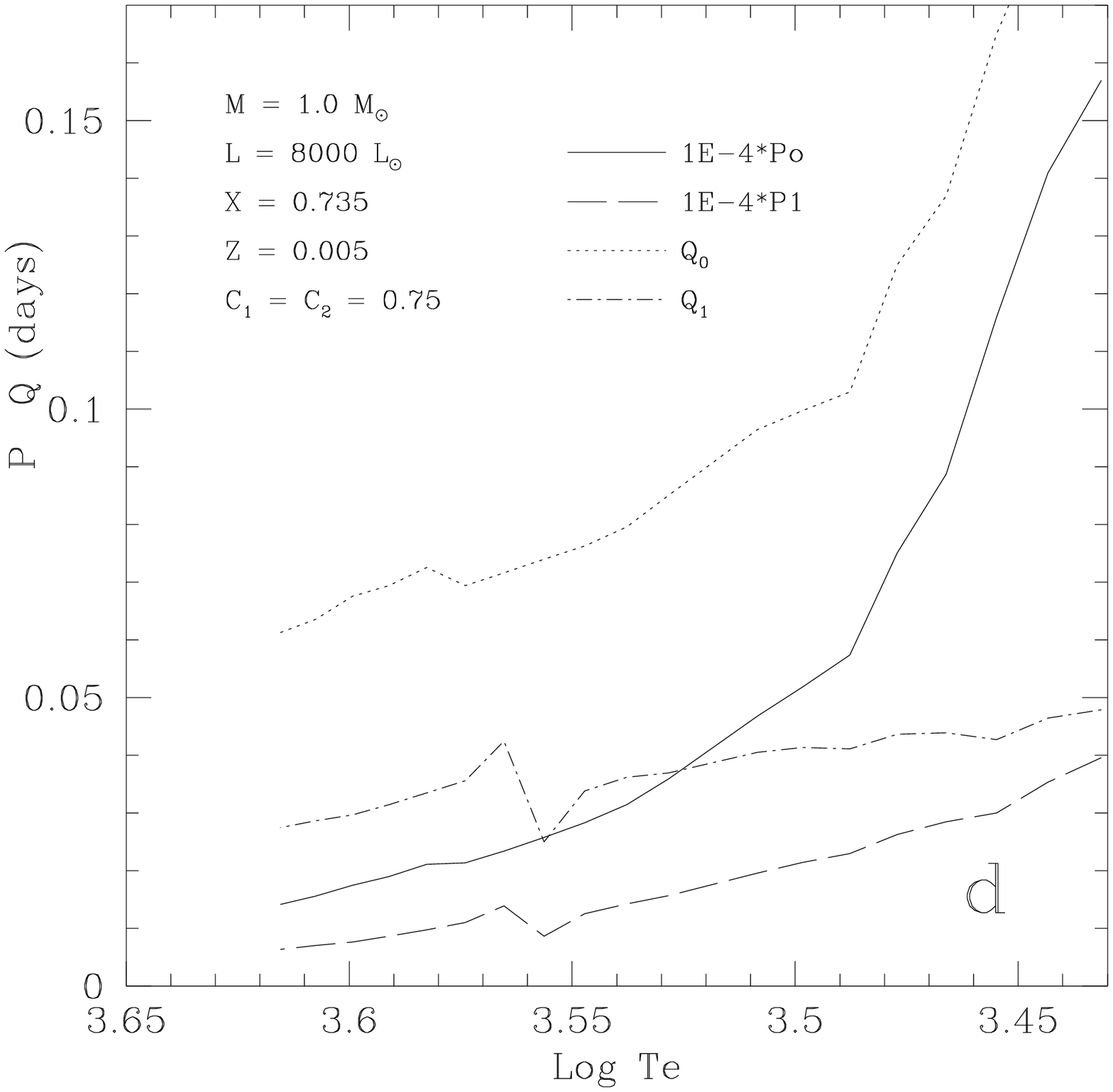,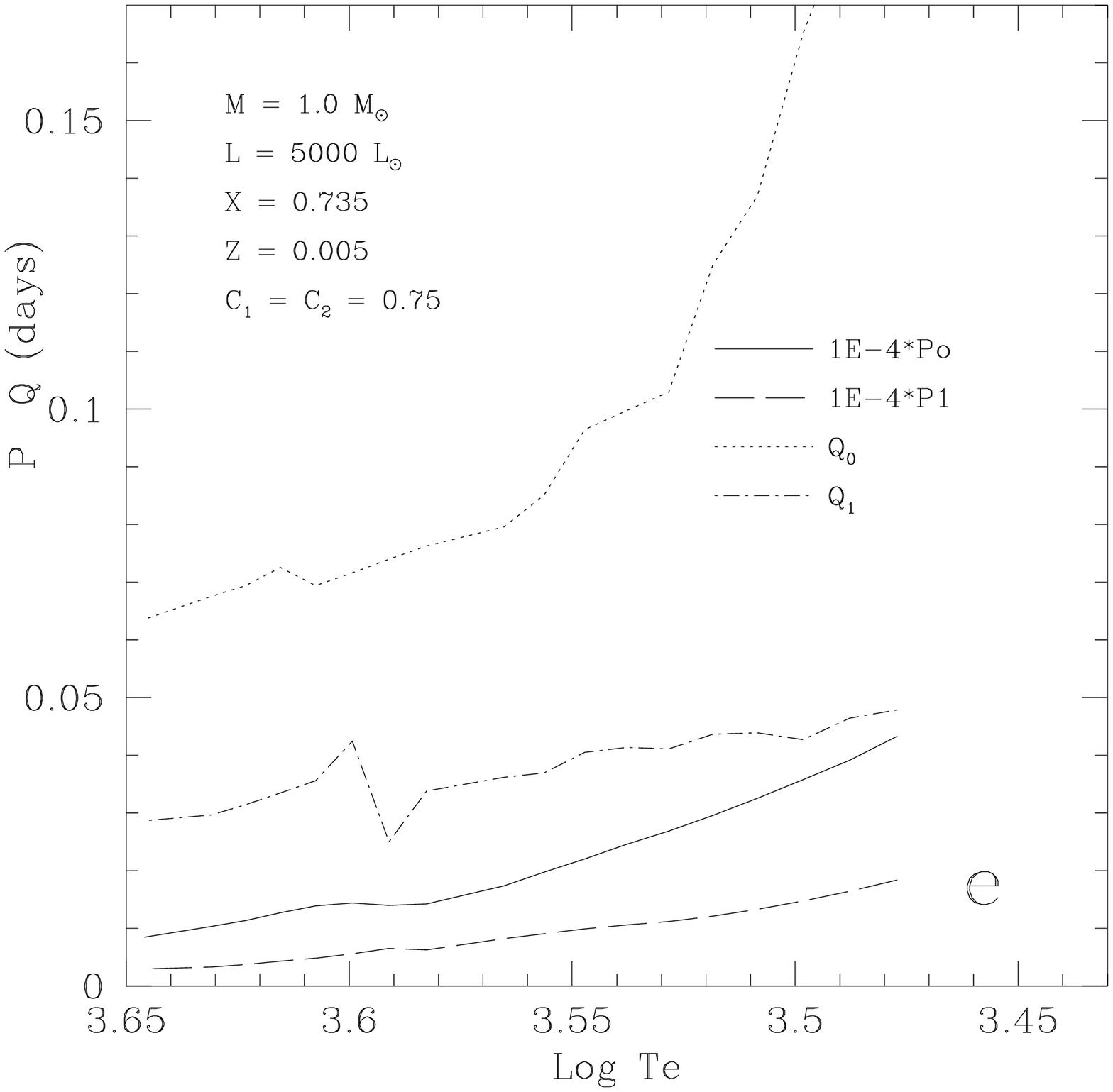,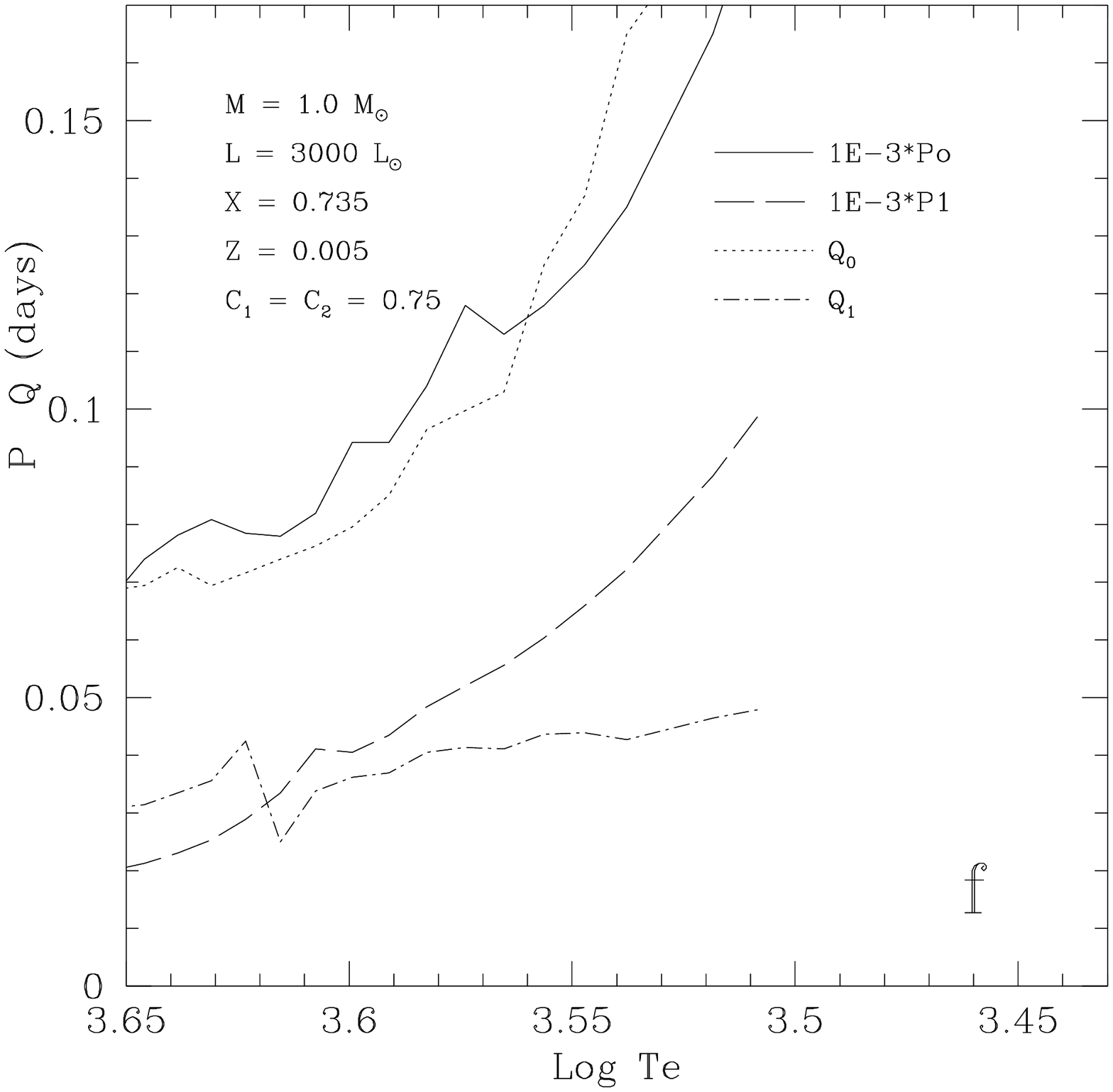]{
The variations of the pulsation periods P and the pulsation constants Q
for the fundamental mode and the first overtone, against the effective temperature.
The coupling between convection and oscillations is included.
\label{fig5}}

\newpage
\setcounter{figure}{0}
\begin{figure}
\centerline{\psfig{figure=fig1a.ps,width=7cm}\psfig{figure=fig1d.ps,width=7cm}}
\centerline{\psfig{figure=fig1b.ps,width=7cm}\psfig{figure=fig1e.ps,width=7cm}}
\centerline{\psfig{figure=fig1c.ps,width=7cm}\psfig{figure=fig1f.ps,width=7cm}}
\caption{ }
\end{figure}

\begin{figure}
\centerline{\psfig{figure=fig2a.ps,width=7cm}}
\centerline{\psfig{figure=fig2b.ps,width=7cm}}
\centerline{\psfig{figure=fig2c.ps,width=7cm}}
\caption{ }
\end{figure}

\begin{figure}
\centerline{\psfig{figure=fig3a.ps,width=7cm}}
\centerline{\psfig{figure=fig3b.ps,width=7cm}}
\centerline{\psfig{figure=fig3c.ps,width=7cm}}
\caption{ }
\end{figure}

\begin{figure}
\centerline{\psfig{figure=fig4a.ps,width=7cm}}
\centerline{\psfig{figure=fig4b.ps,width=7cm}}
\centerline{\psfig{figure=fig4c.ps,width=7cm}}
\caption{ }
\end{figure}

\begin{figure}
\centerline{\psfig{figure=fig5a.ps,width=7cm}\psfig{figure=fig5d.ps,width=7cm}}
\centerline{\psfig{figure=fig5b.ps,width=7cm}\psfig{figure=fig5e.ps,width=7cm}}
\centerline{\psfig{figure=fig5c.ps,width=7cm}\psfig{figure=fig5f.ps,width=7cm}}
\caption{ }
\end{figure}

\begin{figure}
\centerline{\psfig{figure=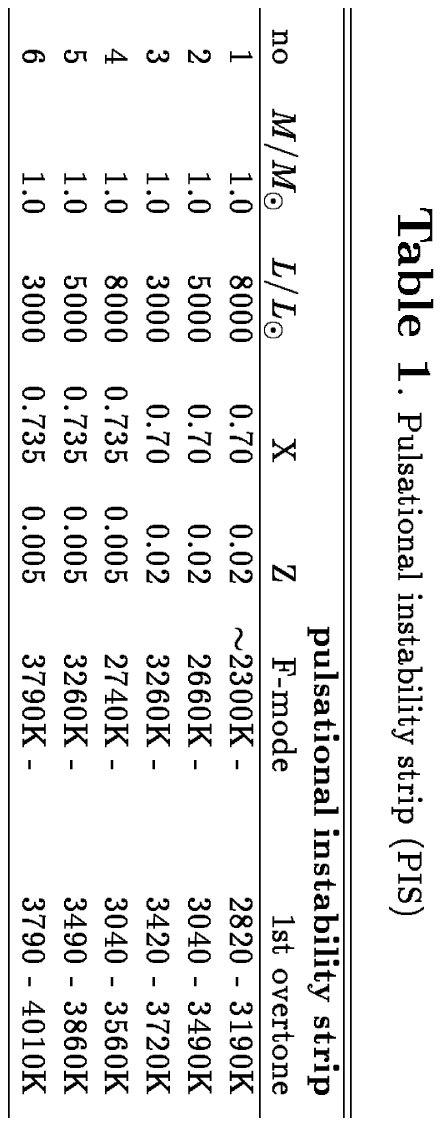}}
\end{figure}
\begin{figure}
\centerline{\psfig{figure=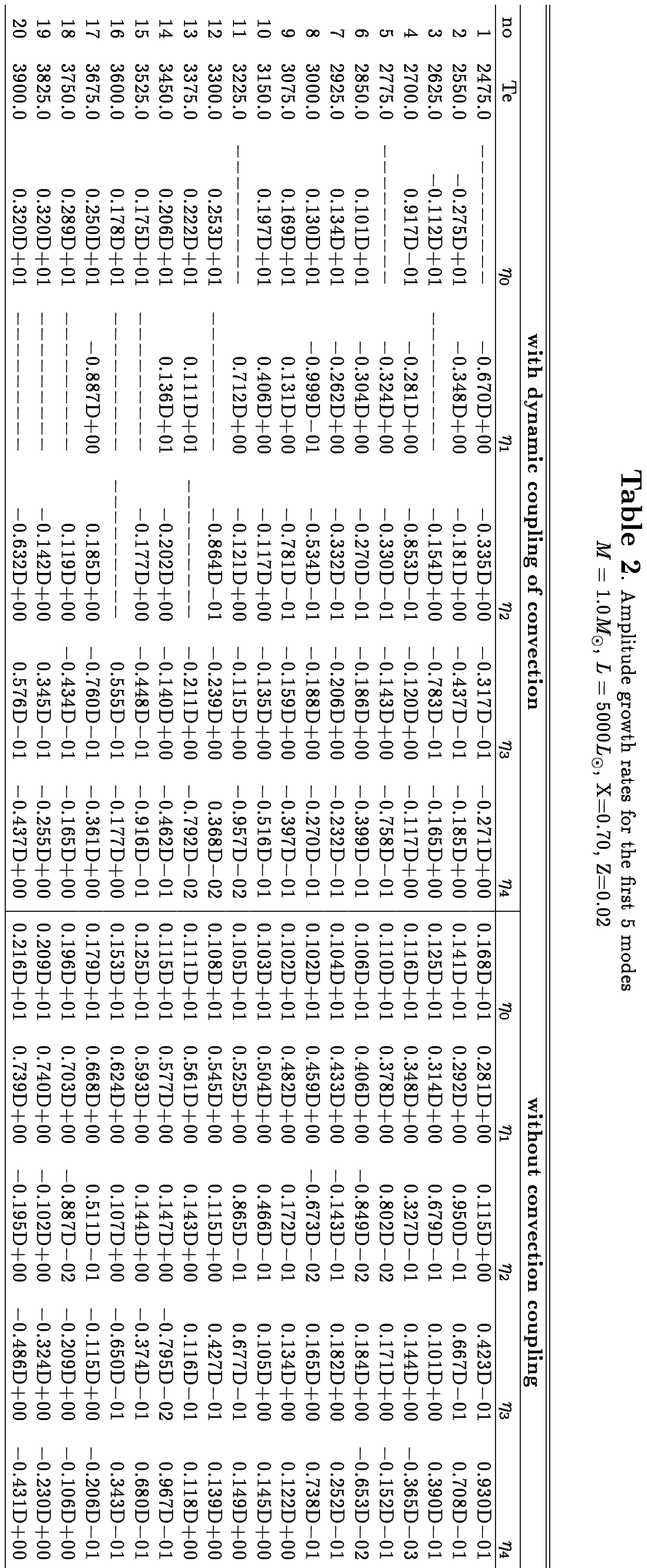}}
\end{figure}
\begin{figure}
\centerline{\psfig{figure=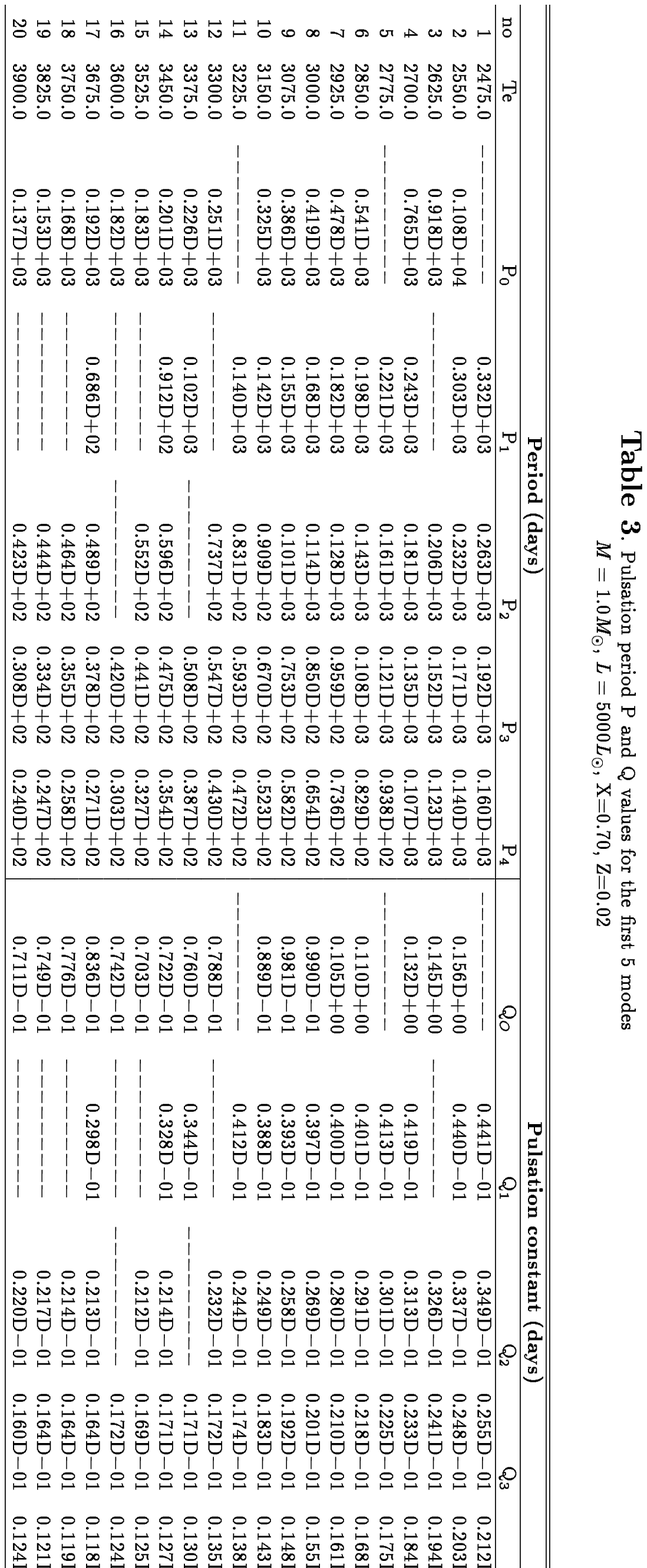}}
\end{figure}

\end{document}